\documentclass[journal=jacsat,manuscript=article]{achemso}

\usepackage[version=3]{mhchem} 
\usepackage{graphicx} %
\usepackage{float} %
\usepackage{subfigure} %
\usepackage{svg}
\usepackage{multirow}
\usepackage{multicol}
\usepackage{booktabs}
\usepackage[utf8]{inputenc}
\usepackage{diagbox}
\usepackage{textcomp} 
\usepackage{gensymb}  
\usepackage{longtable}
\usepackage{caption}
\usepackage{subcaption}
\usepackage{multirow}
\usepackage{xr}
\newcommand*{\citen}[1]{%
  \begingroup
    \romannumeral-`\x 
    \setcitestyle{numbers}%
    \cite{#1}%
  \endgroup   
}
\author{Ziwei Chai}
\email{ziwei.chai@chem.uzh.ch}
\author{Sandra Luber}
\affiliation[University of Zurich]
{Department of Chemistry, University of Zurich, Winterthurerstrasse 190, 8057 Zurich, Switzerland}

\title[An \textsf{achemso} demo]
  {Stable, Fast, and Accurate Kohn-Sham Inversion in Gaussian Basis for Open Shell Molecular and Condensed Phase Systems via Density Matrix Penalization} 

\keywords{American Chemical Society, \LaTeX}

\begin{document}

\begin{tocentry}
\includegraphics{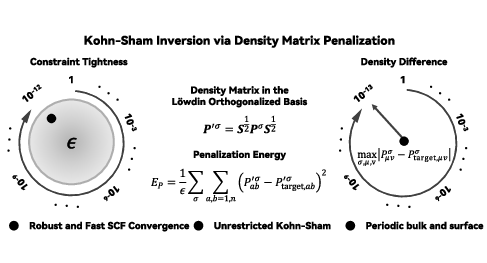}
\quad
\label{For Table of Contents Only}
\end{tocentry}

\begin{abstract}
Here we present a  density matrix based KS inversion method formulated entirely within a Gaussian basis representation to optimize a KS potential matrix that reproduces a target electron density. Inverse Kohn–Sham (KS) density functional theory (DFT) aims to determine the effective local KS potential that reproduces a target electron density, and is important both for electronic structure analysis and for the development of orbital based correction methods. In finite Gaussian basis implementations, however, conventional inverse KS-DFT approaches such as the Zhao–Morrison–Parr (ZMP) method often become poorly constrained and inefficient, because the real space penalty potential is projected onto a limited number of Gaussian basis matrix elements, which can strongly coarse-grain its spatial variation. In the present method, the density matrix mismatch is defined in a Löwdin orthogonalized basis, which yields a penalty energy invariant under unitary rotations in that basis. The corresponding penalty potential contribution to the KS Hamiltonian is derived analytically in the original nonorthogonal Gaussian basis. Across a wide range of penalty strengths, the self consistent field (SCF) optimization remains robust and efficient for various open shell systems, while progressively tightening the penalty drives the electron density into accurate agreement with the target. Benchmarks on molecules and condensed phase systems show that the method achieves substantially smaller attainable density deviations than the conventional ZMP method. The method provides a fast and accurate route to KS inversion in finite Gaussian basis sets and may also be useful for future orbital based correction schemes.\\
\end{abstract}

\section{1. Introduction}
Given a target ground state electron density, one may seek an effective KS description whose self consistent density reproduces that target as closely as possible.\cite{HK_pr_1964, ks_pr_1965, levy_pnas_1979, Leeuwen_Baerends_pra_1994, ZMP_pra_1994, Yang_Wu_prl_2002, Wu_Yang_jcp_2003, Zhang_Carter_jcp_2018, Tim_jcp_2023, Shi_Adam_jpcl_2021, Jensen_ijqc_2018}. When the target density is obtained from a higher level electronic structure method, such as CCSD(T),\cite{ccsd_jcp_1982, ccsdt_cpl_1989} DMRG,\cite{dmrg_prl_1992, dmrg_prb_1993, dmrg_jcp_1999} or QMC,\cite{qmc_prl_1980, qmc_rmp_2001} the resulting KS representation provides an effective one particle description of the many body effects encoded in that density.\cite{Erhard_jcp_2022, Knorr_prl_1992, Knorr_prb_1994, Wagner_prb_2014} Such density constrained KS reconstructions, including formulations commonly referred to as inverse KS-DFT, are useful for analyzing the performance of approximate exchange-correlation (XC) approximations, guiding functional development,\cite{Leeuwen_Baerends_pra_1994, Shi_Adam_jpcl_2021, Erhard_jcp_2022} and generating reference data for machine-learned XC models.\cite{Zhuang_jacsau_2025}\\
Accordingly, substantial efforts have been devoted to inverse KS-DFT and related density-to-potential reconstruction methods\cite{Aryasetiawan_prb_1988, Knorr_prl_1992, Gorling_pra_1992, Zhao_Parr_jcp_1993, Wang_Parr_pra_1993, ZMP_pra_1994, Knorr_prb_1994, Leeuwen_Baerends_pra_1994, Yang_Wu_prl_2002, Wu_Yang_jcp_2003, Peirs_pra_2003, Kadantsev_pra_2004, Bulat_jcp_2007, Gaiduk_jctc_2013, Wagner_prb_2014, Kvaal_jcp_2014, Jensen_ijqc_2018, Zhang_Carter_jcp_2018, Ou_Carter_JCTC_2018, Laestadius_jcp_2018, Kumar_2019, Kanungo_Vikram_nc_2019, Laestadius_jctc_2019, Penz_prl_2019, Kumar_ijqc_2020, Garrick_prx_2020, Callow_jcp_2020, Penz_prl_2020, Shi_Adam_jpcl_2021, Kumar_cpl_2021, Kanungo_jpcl_2021, Erhard_jcp_2022, Bousiadi_pccp_2022, Tim_jcp_2023, Aouina_prb_2023, Penz_es_2023, Kanungo_jpcl_2023, Trushin_pra_2024, Ravindran_prb_2025, Herbst_prb_2025, Erhard_jcp_2025, Kaiser_jcp_2025}. Several software packages dedicated to inverse DFT are also available, including KS-pies,\cite{nam_jcp_2021} n2v,\cite{shi_wirecms_2022} Serenity,\cite{niemeyer_wirecms_2023} and invDFT.\cite{invDFT_arxiv_2025} Among the proposed approaches, one important class employs a tunable scaling parameter to control the strength of the penalty potential that enforces real space electron density matching\cite{Zhao_Parr_jcp_1993, Wang_Parr_pra_1993, ZMP_pra_1994, Kumar_ijqc_2020, Penz_es_2023, Herbst_prb_2025}. The Zhao-Morrison-Parr (ZMP) method\cite{Zhao_Parr_jcp_1993, Wang_Parr_pra_1993, ZMP_pra_1994} constructs a Coulomb-like penalty potential from the difference between the current and target real space electron densities, scales it by a penalty parameter $\lambda$, and adds it to the KS Hamiltonian for a KS-SCF optimization. Repeating the above procedure with increasingly larger $\lambda$ pushes the self consistent real space electron density toward the target. Recent studies have shown that the above procedure can be understood, at the level of the abstract density-potential inversion problem, as an instance of the Moreau-Yosida (MY) regularization framework\cite{Penz_es_2023, Kvaal_jcp_2014, Laestadius_jcp_2018, Laestadius_jctc_2019, Penz_prl_2019, Penz_prl_2020}. In this interpretation, the inversion problem remains differentiable for finite constraining strength ($\epsilon = 1/\lambda$, with larger $\lambda$ corresponding to a stronger constraint), while the exact inversion is approached in the limit $\epsilon \to 0$ (equivalently, $\lambda \to \infty$). Recently, Herbst et al. demonstrated that, within a plane wave basis representation (adopted in the DFTK software package), by decreasing the regularization parameter $\epsilon$ to about $1.0\times10^{-7}$ (equivalently, increasing $\lambda$ to about $1.0\times10^{7}$), the XC potential given by a forward PBE ground state calculation for periodic systems can be accurately recovered by using the exact MY regularized formulation\cite{Herbst_prb_2025}.\\
Many advances have been made toward accurate and efficient inverse KS-DFT and related local potential construction methods, particularly for calculations based on local basis sets.\cite{Bulat_jcp_2007, Heaton_jcp_2008, Jacob_jcp_2011, Gaiduk_jctc_2013, Ou_Carter_JCTC_2018, Kanungo_Vikram_nc_2019, Callow_jcp_2020, Kumar_cpl_2021, Trushin_jcp_2021, Bousiadi_pccp_2022, Erhard_jcp_2022, Oueis_jctc_2022, Tim_jcp_2023, Erhard_jcp_2025} Despite there are these progresses, inverse KS-DFT and related local potential construction schemes remain numerically challenging in calculations based on local basis sets, especially finite Gaussian basis sets.\cite{Schipper_tca_1997, Mura_jcp_1997, Staroverov_jcp_2006, Jacob_jcp_2011, Gaiduk_jctc_2013, Mayer_jctc_2017, Jensen_ijqc_2018, Shi_Adam_jpcl_2021, Tim_jcp_2023, Trushin_pra_2024, Kaiser_jcp_2025} However, within KS-DFT formulated in a finite Gaussian basis, constructing a penalty potential matrix from differences in the real space density inevitably introduces a mismatch between the real space grid and the finite Gaussian representation. In particular, fine spatial variations of a real space penalty potential are ultimately compressed into a finite set of matrix elements through integration with products of basis functions. Such severe coarse grain substantially weakens the ability of the constraining potential to adjust the electron density, and in practice leads to convergence difficulties, plateaus in density matching accuracy, and severe slowdowns of the overall optimization. These numerical issues typically become more pronounced as the regularization parameter $\epsilon$ is small, and thus constitute a bottleneck for the successful application of this approach. By contrast, within a Gaussian basis representation the more reasonable constrained quantity is the electron density matrix. It can be obtained directly and accurately by solving the KS equations for the corresponding matrix form KS potential and filling the resulting KS orbitals, and it generates the real space density through the basis functions. It is therefore desirable to develop an inverse KS-DFT scheme that is consistently defined in a Gaussian basis representation, takes the density matrix as the constrained variable, and inherits the practical differentiability and exactness at $\epsilon \to 0$ of the ZMP method or the MY regularization framework.\\
This choice is also motivated by a basic conceptual limitation of inverse KS-DFT in finite Gaussian basis sets.\cite{Jacob_jcp_2011, desilva_2012_pra, Gaiduk_jctc_2013, Shi_Adam_jpcl_2021} In the complete basis limit, the exact ground state density determines the local XC potential up to an additive constant.\cite{HK_pr_1964, ks_pr_1965} In a finite Gaussian basis, however, the inversion can determine only a projected potential within a restricted representation, and this projected object is generally not unique unless an additional selection criterion is imposed.\cite{Jacob_jcp_2011, Gaiduk_jctc_2013, Shi_Adam_jpcl_2021} Consequently, within a finite Gaussian basis, elementwise agreement between the XC potential matrix from a forward KS calculation and a constraining potential matrix reconstructed from the target density is, in general, neither guaranteed nor by itself a meaningful validation criterion. In this work, we therefore focus instead on reproducibility of the target density, or equivalently of the target density matrix in the chosen representation, as the quantity that is both well defined and directly assessable in the finite basis setting. A more systematic investigation of potential consistency and its connection to specific regularization choices is an interesting topic, which we leave for future work.\cite{Bulat_jcp_2007, Heaton_jcp_2008, Callow_jcp_2020, Trushin_jcp_2021}\\
Motivated by these considerations, we develop a density matrix penalization inverse KS-DFT scheme for Gaussian basis KS-DFT. The penalty energy is defined in terms of the Löwdin transformed target and current density matrices. This choice is essential because, in the nonorthogonal Gaussian atomic orbital basis, a penalty constructed from direct elementwise differences of raw density matrix elements would be basis dependent. Specifically, the penalty energy is defined as the sum of squared elementwise deviations between the transformed density matrices, scaled by $1/\epsilon$. The corresponding penalty potential matrix in the original basis is derived directly by differentiating the penalty energy with respect to the density matrix, which preserves tensor consistency between the penalty definition and the resulting potential matrix. With the penalty energy and the corresponding penalty potential defined in this way, the method follows the ZMP strategy (equivalently, the MY regularization) of monotonically increasing the scaling parameter $1/\epsilon$ to drive the constraining strength and the matching accuracy toward their limiting values.\\
We implemented the method and also the conventional ZMP approach based on real space density differences, suitable for spin unrestricted KS framework, in the CP2K software package. We applied the proposed method to a series of open shell systems and compared the results with those obtained from the conventional ZMP approach. In terms of the achievable accuracy of the target density, the density matrix penalization method achieves a maximum real space electron density deviation as low as about $1.0\times10^{-13}$-$1.0\times10^{-12}$ for most systems over all grid points. By comparison, this minimum attainable deviation is approximately seven orders of magnitude smaller than that obtained with the ZMP method. In addition, numerical tests show that for most systems the proposed method maintains robust convergence and a relatively small number (less than 1000) of SCF iterations over a wide range of penalty parameters, with $\epsilon$ varying from $1.0\times10^{0}$ down to $1.0\times10^{-10}$ or $1.0\times10^{-11}$. In contrast, the ZMP method fails to achieve SCF convergence once  $\epsilon < 1.0\times10^{-4}$ for all of the test systems. Moreover, the number of SCF iterations required for ZMP to reach convergence increases rapidly as $\epsilon$ decreases and becomes significantly larger than that required by the proposed density matrix penalization method.\\
The remainder of this paper is organized as follows. Section 2 presents theoretical formulation and the algorithmic workflow. Section 3 describes the computational setups and test models. Section 4 shows the test result for the proposed method and the ZMP methods, including: the reproduction accuracy of target electron densities of the proposed method (Subsection 4.1), the comparison with the ZMP method in terms of the attainable limit of accuracy (Subsection 4.2), and comparing the computational efficiency and convergence of the proposed method and the ZMP method (Subsection 4.3). We summarize and conclude our work in Section 5.\\
\section{2. Method}
\subsection{2.1 Background: KS equation on a Gaussian basis set}
We consider a simulation box containing a set of Gaussian basis functions $\phi_a(\boldsymbol{r}) \quad (a=1,\dots,n)$. $n$ denotes the number of Gaussian basis functions in the simulation cell. The overlap integral between $\phi_a(\boldsymbol{r})$ and $\phi_b(\boldsymbol{r})$ is $S_{a b}$. So, the corresponding matrix $\boldsymbol{S}$ here is the overlap matrix. The matrix formulation of the KS equations for spin channel $\sigma$ is
\begin{equation}
\boldsymbol{K}^\sigma \boldsymbol{C}^\sigma=\boldsymbol{S} \boldsymbol{C}^\sigma \boldsymbol{E}^\sigma.
\label{eq1}
\end{equation}
$\boldsymbol{K}^\sigma$ is the KS Hamiltonian matrix. $\boldsymbol{C}^\sigma$ is the matrix of KS orbital expansion coefficients in the Gaussian basis where the $i$th column contains the expansion coefficients of the $i$th KS orbital $\psi_i^\sigma(\boldsymbol{r})\left(\psi_i^\sigma(\boldsymbol{r})=\sum_{a=1, n} C_{a i}^\sigma \phi_a(\boldsymbol{r})\right)$. $ \boldsymbol{E}^\sigma$ is the diagonal matrix of KS orbital eigenvalues, with $E_{i i}=\epsilon_i$ corresponding to the $i$th KS orbital and $E_{i j}=0$ for $i \neq j$. In this work, we present the formulation for real-valued KS orbitals. The generalization to complex orbitals is straightforward. $f_i$ denotes the occupation number of KS orbital $i$, and the electron density then can be calculated by using
\begin{equation}
\begin{aligned}
& n^\sigma(\boldsymbol{r})=\sum_{i=1, m} f_i^\sigma\left(\psi_i^\sigma(\boldsymbol{r})\right)^2=\sum_{i=1, m} \sum_{a=1, n} \sum_{b=1, n} f_i^\sigma C_{a i}^\sigma C_{b i}^\sigma \phi_a(\boldsymbol{r}) \phi_b(\boldsymbol{r}) \\
& \quad=\sum_{a=1, n} \sum_{b=1, n} \phi_a(\boldsymbol{r}) \phi_b(\boldsymbol{r}) \sum_{i=1, m} f_i^\sigma C_{a i}^\sigma C_{b i}^\sigma=\sum_{a=1, n} \sum_{b=1, n} \phi_a(\boldsymbol{r}) \phi_b(\boldsymbol{r}) P_{a b}^\sigma .
\end{aligned}
\label{eq2}
\end{equation}
$m$ denotes the number of KS orbitals in spin channel $\sigma$. Here, the density matrix for spin channel $\sigma$ is defined as
\begin{equation}
P_{a b}^\sigma=\sum_{i=1, m} f_i^\sigma C_{a i}^\sigma C_{b i}^\sigma.
\label{eq3}
\end{equation}\\
\subsection{2.2 Rotational invariant penalty energy}
The penalty energy defined from the deviation between the current and target density matrices should not depend on a particular choice of the basis used to represent the subspace spanned by the original Gaussian basis. Here, we consider the density matrix represented in the Löwdin orthonormalized basis. From Eq. (1), we have
\begin{equation}
\begin{aligned}
\boldsymbol{S}^{-\frac{1}{2}} \boldsymbol{K}^\sigma \boldsymbol{S}^{-\frac{1}{2}} \boldsymbol{S}^{\frac{1}{2}} \boldsymbol{C}^\sigma & =\boldsymbol{S}^{\frac{1}{2}} \boldsymbol{C}^\sigma \boldsymbol{E}^\sigma \\
\boldsymbol{K}^{\prime \sigma} \boldsymbol{C}^{\prime \sigma} & =\boldsymbol{C}^{\prime \sigma} \boldsymbol{E}^\sigma,
\end{aligned}
\label{eq4}
\end{equation}
in which $\boldsymbol{K}^{\prime \sigma}=\boldsymbol{S}^{-\frac{1}{2}} \boldsymbol{K}^\sigma \boldsymbol{S}^{-\frac{1}{2}}$ and $\boldsymbol{C}^{\prime \sigma} \boldsymbol{=} \boldsymbol{S}^{\frac{1}{2}} \boldsymbol{C}^{\sigma}$ are KS Hamiltonian matrix and coefficient matrix represented in the Löwdin orthonormalized basis. The density matrix $\boldsymbol{P}^{\prime \sigma} $ represented on this basis is given by
\begin{equation}
 \boldsymbol{P}^{\prime \sigma}= \boldsymbol{C}^{\prime \sigma} \boldsymbol{f}^{\sigma}  \boldsymbol{C}^{\prime \sigma^T}= \boldsymbol{S}^{\frac{1}{2}}  \boldsymbol{C}^\sigma  \boldsymbol{f}^{\sigma}\boldsymbol{C}^{\sigma T}  \boldsymbol{S}^{\frac{1}{2}}= \boldsymbol{S}^{\frac{1}{2}}  \boldsymbol{P}^\sigma  \boldsymbol{S}^{\frac{1}{2}} ,
\label{eq5}
\end{equation}
where \(\boldsymbol{f}^{\sigma}\) is the diagonal matrix of orbital occupation numbers $f_i^\sigma$ for spin \(\sigma\). We formulate the penalty energy as the sum of the squared element-wise differences between the current SCF step $\boldsymbol{P}^{\prime \sigma}$ and the target $\boldsymbol{P}_{\text {target }}^{\prime \sigma} $ divided by the regularization parameter $\epsilon$
\begin{equation}
E_P=\frac{1}{\epsilon} \sum_\sigma E_P^\sigma=\frac{1}{\epsilon} \sum_\sigma \sum_{a, b=1, n}\left(P_{a b}^{\prime \sigma}-P_{t a r g e t, ab}^{\prime\sigma}\right)^2=\frac{1}{\epsilon} \sum_\sigma \operatorname{Tr}\left[\Delta{\boldsymbol{P}^{\prime}}^\sigma \Delta{\boldsymbol{P}^{\prime}}^\sigma\right],
\label{eq6}
\end{equation}
where $\Delta \boldsymbol{P}^{\prime \sigma}=\boldsymbol{P}^{\prime \sigma}-\boldsymbol{P}_{\text {target }}^{\prime \sigma}$. 
From Eq.~(6), we define the penalty energy in the Löwdin orthonormalized basis and examine its behavior under an arbitrary orthogonal transformation in that basis. Let $\boldsymbol U$ be an orthogonal matrix acting in the Löwdin orthonormalized basis, such that $\boldsymbol U^T \boldsymbol U = \boldsymbol U \boldsymbol U^T = \boldsymbol I$. Under this transformation, the density matrix difference $\Delta \boldsymbol{P}^{\prime \sigma}$ transforms as $\Delta \boldsymbol{P}^{\prime \sigma} \rightarrow \boldsymbol U^T \Delta \boldsymbol{P}^{\prime \sigma} \boldsymbol U$. The penalty energy is then
\begin{equation}
E_P=\frac{1}{\epsilon} \sum_\sigma \operatorname{Tr}\left[
\boldsymbol{U}^T \Delta \boldsymbol{P}^{\prime \sigma} \boldsymbol{U}\,
\boldsymbol{U}^T \Delta \boldsymbol{P}^{\prime \sigma} \boldsymbol{U}
\right]
=\frac{1}{\epsilon} \sum_\sigma \operatorname{Tr}\left[
\boldsymbol{U}^T \Delta \boldsymbol{P}^{\prime \sigma}
\Delta \boldsymbol{P}^{\prime \sigma} \boldsymbol{U}
\right]
=\frac{1}{\epsilon} \sum_\sigma \operatorname{Tr}\left[
\Delta \boldsymbol{P}^{\prime \sigma}
\Delta \boldsymbol{P}^{\prime \sigma}
\right],
\label{eq7}
\end{equation}
where the last equality follows from the cyclic invariance of the trace together with $\boldsymbol U \boldsymbol U^T=\boldsymbol I$. Therefore, $E_P$ is invariant under orthogonal transformations of the Löwdin-orthonormalized basis. The KS-DFT total energy then becomes
\begin{equation}
E_{\mathrm{tot}} = E_{\mathrm{0}} + E_P ,
\label{eq8}
\end{equation}
where $E_{0}$ denotes the DFT total energy functional excluding the XC energy, and $E_P$ is the penalty energy introduced to enforce agreement with the target density matrix.\\
\subsection{2.3 Penalty potential matrix in the KS Hamiltonian matrix}
By minimizing the energy functional in Eq. (8), one can at the end obtain a density matrix around the target by solving the resulting KS equation. The equation needs to be solved self-consistently and the analytical form of the Hamiltonian matrix is derived from the total energy expression in Eq. (8). The contribution to the Hamiltonian by $E_{\text {0 }}$ is just the standard KS-DFT Hamiltonian $\boldsymbol{K}_{\text {0 }}^{\sigma}$, except that in the inverse DFT calculation, the Hartree potential is replaced by the one generated from the target electron density and the XC potential is always set to zero. The potential contribution of $E_P$ given in Eq. (6) to the KS Hamiltonian matrix $\boldsymbol{K}^\sigma$ can be derived as follows (the Einstein summation convention\cite{einstein1916foundation} is applied here, in which repeated indices inside the same expression are summed over unless the indices are enclosed in parentheses)
\begin{equation}
\begin{gathered}
\left(K_P^{(\sigma)}\right)_{c d}
=
\frac{\partial}{\partial P_{dc}^{(\sigma)}}
\left[
\frac{1}{\epsilon}
\sum_{\sigma'}
\sum_{a,b}
\left(
P_{(a)(b)}^{\prime(\sigma')}
-
P_{t a r g e t,(a)(b)}^{\prime(\sigma')}
\right)^2
\right]
\\
=
\frac{1}{\epsilon}
\sum_{a,b}
\frac{\partial
\left(
P_{(a)(b)}^{\prime(\sigma)}
-
P_{t a r g e t,(a)(b)}^{\prime(\sigma)}
\right)^2
}
{\partial P_{ef}^{\prime(\sigma)}}
\frac{\partial P_{ef}^{\prime(\sigma)}}{\partial P_{dc}^{(\sigma)}}
\\
=
\frac{2}{\epsilon}
\left(
P_{ef}^{\prime(\sigma)}
-
P_{t a r g e t, ef}^{\prime(\sigma)}
\right)
\frac{\partial
\left(
\left(S^{\frac{1}{2}}\right)_{ei}
P_{ik}^{(\sigma)}
\left(S^{\frac{1}{2}}\right)_{kf}
\right)}
{\partial P_{dc}^{(\sigma)}}
\\
=
\frac{2}{\epsilon}
\left(
P_{ef}^{\prime(\sigma)}
-
P_{t a r g e t, ef}^{\prime(\sigma)}
\right)
\left(S^{\frac{1}{2}}\right)_{ei}
\delta_{id}
\delta_{ck}
\left(S^{\frac{1}{2}}\right)_{kf}
\\
=
\frac{2}{\epsilon}
\left(S^{\frac{1}{2}}\right)_{cf}
\Delta P_{fe}^{\prime(\sigma)T}
\left(S^{\frac{1}{2}}\right)_{ed}
\\
=
\frac{2}{\epsilon}
\left(
\boldsymbol{S}^{\frac{1}{2}}
\left(\Delta \boldsymbol{P}^{\prime(\sigma)}\right)^T
\boldsymbol{S}^{\frac{1}{2}}
\right)_{cd}.
\end{gathered}
\label{eq9}
\end{equation}
For each spin channel $\sigma$, matrix $\boldsymbol{K}_P^\sigma$ can be directly added with $\boldsymbol{K}_{\text {0 }}^{\sigma}$ to obtain the KS Hamiltonian matrix $\boldsymbol{K}^\sigma=\boldsymbol{K}_{\text {0 }}^\sigma+\boldsymbol{K}_P^\sigma $. $\epsilon$ controls the tightness of the constraint, with smaller $\epsilon$ leading to a tighter constraint to the electron density.
\subsection{2.4 Optimization Procedure}
\begin{figure}[H]
\centering
\includegraphics[width=0.5\textwidth]{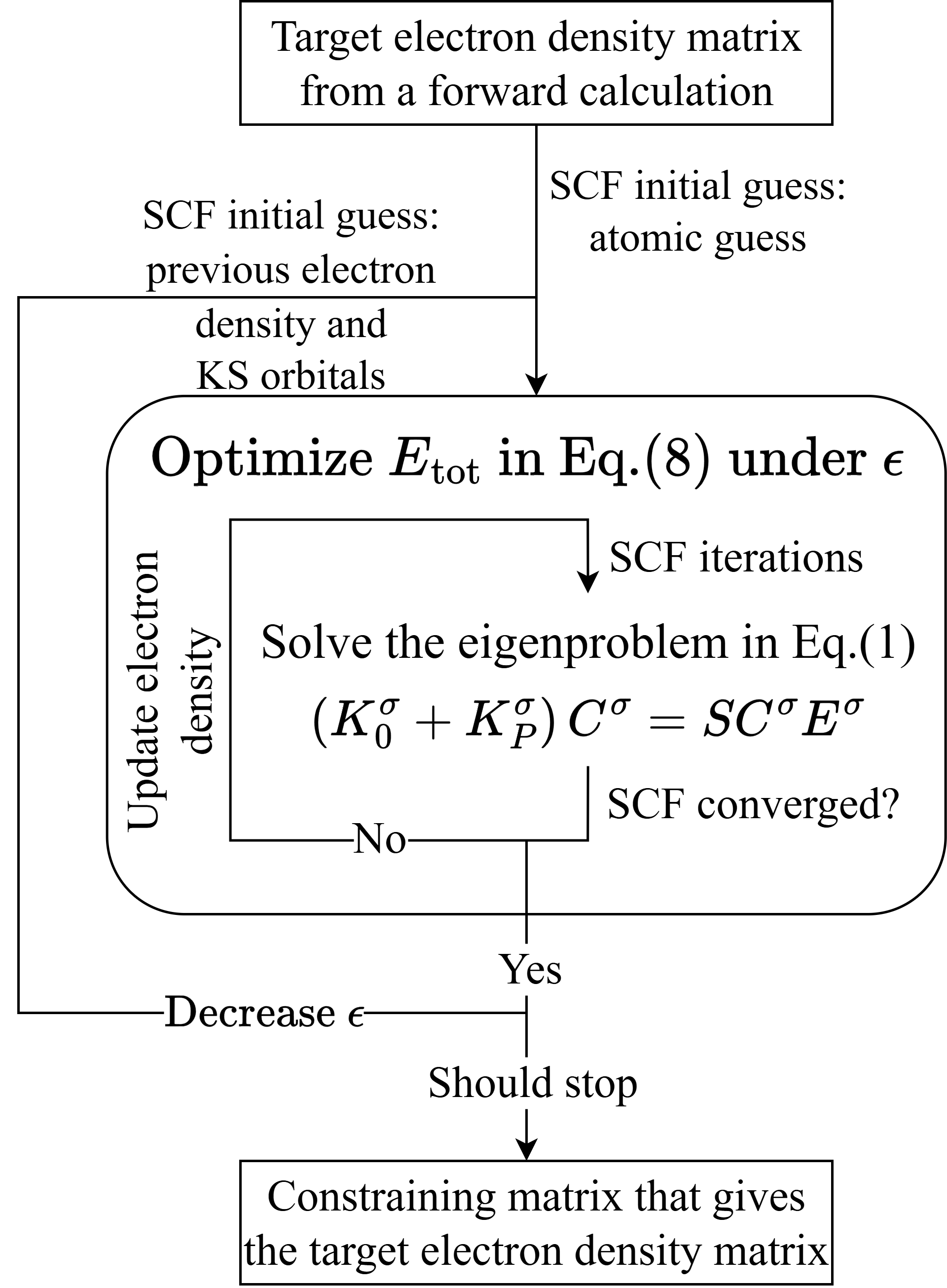}
\caption{Schematic illustration of the penalty strength iteration and KS-DFT SCF optimization procedure. The inner loop performs SCF iterations at fixed $\epsilon$ values, using an initial guess from either atomic calculations or the previously converged electron density and KS orbitals. The outer loop gradually decreases $\epsilon$ until satisfactory density matrix agreement is achieved.}\label{fig1}
\end{figure}

The penalty strength must be sufficiently large to ensure that the deviation between the converged SCF density matrix and the target density matrix is sufficiently small. In practice, we follow the strategy proposed in Ref.~\citen{Herbst_prb_2025}, in which one SCF optimization is performed at a fixed value of $\epsilon$, and, once SCF convergence is achieved or the maximum number of SCF iterations is reached, $\epsilon$ is reduced and the next SCF optimization is carried out. In the first SCF optimization of an inverse DFT calculation, with $\epsilon=1$, the atomic guess is used as the initial guess for the electron density and KS orbitals. After each SCF optimization, if the SCF procedure has either converged or reached the maximum iteration limit of 6000 steps, the parameter $\epsilon$ is reduced by one order of magnitude, and a new SCF optimization is then performed using the updated value of $\epsilon$. In our calculations, $\epsilon$ was reduced from 1 until $\epsilon=1.0\times10^{-12}$ ($1\to0.1\to0.01\to\cdots\to1\times10^{-12}$). Numerical tests showed that this strategy is sufficiently mild to ensure stable SCF convergence. A schematic illustration of the overall inverse DFT procedure is shown in Fig.~\ref{fig1}.
\section{3. Computational Setups and Models}
\subsection{3.1 Computational Setups}
All the calculations were performed using our modified version of the Quickstep module of the CP2K software package\cite{quickstep_cpc_2005,jpca_cp2k_luber}. In Quickstep, the KS matrix is represented and solved in a Gaussian basis, whereas the Hartree potential is calculated in reciprocal space and transferred back to real space to be integrated with the Gaussian basis in the construction of the KS matrix\cite{quickstep_cpc_2005, quickstep_prev}. The nuclei together with the core nonvalence electrons were treated effectively by means of norm-conserving Goedecker-Teter-Hutter pseudopotentials\cite{gth1, gth2, gth3}. The standard MOLOPT sets, optimized for molecular systems, were used for molecules, while the MOLOPT-SR variants with shorter radial tails were adopted for bulk and surface systems\cite{molopt}. PBE\cite{pbe} was used as the XC functional in geometry optimization and forward KS-DFT calculations, and in some cases it was corrected using Dudarev’s +U formalism\cite{dudarev_u}. In these calculations, the \(3d\) subspaces of \(\mathrm{Cu}\) in \(\mathrm{CuCl}_2\), the \(3d\) subspaces of \(\mathrm{Ti}\) in \(\mathrm{TiO}_2\), the \(3d\) subspaces of \(\mathrm{Ni}\) in \(\mathrm{NiO}\), the \(3d\) subspaces of \(\mathrm{Co}\) in \(\mathrm{CoO}\), and the \(4f\) subspaces of \(\mathrm{Ce}\) in \(\mathrm{CeO}_2\) were corrected by setting \(U-J = 4\,\mathrm{eV}\), which is a physically reasonable correction strength based on our experience\cite{ziwei_plusu_jctc}. In all the inverse DFT calculations, PBE XC contributions were excluded and +U were not activated. In accordance with the PBE functional, the following pseudopotentials were used in all the forward and inverse KS-DFT calculations: GTH-PBE-q6 (O), GTH-PBE-q5 (N), GTH-PBE-q11 (Cu), GTH-PBE-q7 (Cl), GTH-PBE-q4 (C), GTH-PBE-q1 (H), GTH-PBE-q12 (Ti), GTH-PBE-q18 (Ni), GTH-PBE-q17 (Co), GTH-PBE-q12 (Ce), and GTH-PBE-q11 (Ag). The following basis sets were used for molecules: TZVP-MOLOPT-GTH-q6 (O), TZVP-MOLOPT-GTH-q5 (N), TZVP-MOLOPT-PBE-GTH-q11 (Cu), TZVP-MOLOPT-PBE-GTH-q7 (Cl), TZVP-MOLOPT-PBE-GTH-q4 (C), and TZVP-MOLOPT-PBE-GTH-q1 (H). For bulk and surface systems the following basis sets were used: DZVP-MOLOPT-SR-GTH-q12 (Ti), DZVP-MOLOPT-SR-GTH-q6 (O), DZVP-MOLOPT-SR-GTH-q1 (H), DZVP-MOLOPT-SR-GTH-q18 (Ni), DZVP-MOLOPT-SR-GTH-q17 (Co), DZVP-MOLOPT-SR-GTH-q12 (Ce), and DZVP-MOLOPT-SR-GTH-q11 (Ag). All calculations were spin-unrestricted KS calculations, where the KS orbitals in each spin channel were treated separately, with the multiplicity constraints given in detail for each system in Section~3.2. A cutoff of \(320\,\mathrm{Ry}\) was used for the finest grid level in the five-level multigrid setup. A relative cutoff (REL\_CUTOFF) of \(40\,\mathrm{Ry}\) was used to determine the grid on which a Gaussian is mapped. The \(k\)-point sampling was restricted to the \(\Gamma\) point. The total energy was minimized self-consistently using the orbital transformation (OT) method\cite{ot_paper}. The direct inversion in the iterative subspace (DIIS) method was used as the minimizer. The \texttt{FULL\_SINGLE\_INVERSE} preconditioner and the corresponding \texttt{ENERGY\_GAP} of \(0.2\,\mathrm{Hartree}\) were used. In an inverse KS-DFT SCF optimization for a given fixed \(\epsilon\), the parameter (\texttt{ds\_min} variable under the \texttt{qs\_ot\_type} type in the codes) controlling the OT optimization step was manually fixed at \(0.01/\epsilon\) internally in the CP2K program to achieve successful convergence. The SCF convergence criterion of \(1\times10^{-6}\) was used in all forward and inverse KS-DFT calculations to ensure that the derivatives of the total energy with respect to the KS orbital coefficients are sufficiently small. The maximum number of SCF iterations was set to 6000. After the forward KS-DFT SCF optimization, the inverse KS-DFT SCF loops at the first \(\epsilon\) started from the atomic guess of the density matrix and the KS orbitals. Subsequently, all following inverse KS-DFT SCF calculations were initialized from restart guesses based on the previously converged SCF iterations.\\
\subsection{3.2 Structural Models of the Test Systems}
The experimental bond lengths of \(\mathrm{O}_2\) (triplet) and \(\mathrm{NO}\) (doublet) were taken from the NIST Computational Chemistry Comparison and Benchmark Database (CCCBDB)\cite{HuberHerzberg1979, NIST_SRD114_2005, NIST_CCCBDB}. The bond length of the linear molecule \(\mathrm{CuCl}_2\) (doublet) was taken from the range of the results reported by quantum chemical calculations\cite{10.1063/1.1883167}. The Cartesian coordinates of the benzene dimer radical cation (structure D \(({}^{2}B_g)\), \(x\)-displaced, ``half shift'', doublet) were taken from the Supporting Information of Ref.~\citen{C4CP05784H}. The \((4\times2)\) five-layer slab model with the optimized lattice parameters reported in Ref.~\citen{ziwei_plusu_jctc} was used to construct the initial geometries of the rutile \(\mathrm{TiO}_2\)(110) surface with one oxygen vacancy and the rutile \(\mathrm{TiO}_2\)(110) surface with one adsorbed \(\mathrm{OH}\) group. The high-spin (multiplicity \(=3\)) polaron configuration and the adsorption geometry of \(\mathrm{OH}\) (multiplicity \(=2\)) were obtained by relaxing the initial geometries using the Löwdin PBE+U implementation in CP2K until the maximum geometry change and the maximum force component fell below \(1\times10^{-3}\) Bohr and \(1\times10^{-4}\) Hartree/Bohr, respectively. The antiferromagnetic \(\mathrm{NiO}\) and \(\mathrm{CoO}\) (multiplicity \(=1\)) structures were taken from Ref.~\citen{ziwei_plusu_jctc} and from the Materials Project database (material ID: mp-19079)\cite{materials_project}, respectively. The antiferromagnetic orderings of \(\mathrm{NiO}\) and \(\mathrm{CoO}\) were ensured by employing the broken-symmetry (BS) approach for the atomic-orbital occupations assigned in the initialization of the density matrix in CP2K. The initial geometry of bulk \(\mathrm{CeO}_2\) containing one oxygen vacancy was generated based on the \(\mathrm{CeO}_2\) structure from the Materials Project database (material ID: mp-20194)\cite{materials_project}. The high-spin polaron configuration (multiplicity \(=3\)) was then relaxed using the Löwdin PBE+U implementation in CP2K until the maximum geometry change and the maximum force component fell below \(1\times10^{-3}\) Bohr and \(1\times10^{-4}\) Hartree/Bohr, respectively. The structure containing 32 liquid water molecules and one \(\mathrm{Ag}^{2+}\) ion was randomly selected from an equilibrated NVT AIMD trajectory in which the multiplicity of 2 was maintained during the simulation. All of the models of the above-mentioned molecules, bulk systems, and surfaces were placed in a three-dimensional periodic simulation box. These structures are provided in the Supporting Information (Section 6) together with the lattice parameters.\\
\section{4. Results}
In this section, the reproduction accuracy of the target electron densities, computational efficiency, and convergence behavior of the method are tested and reported. We tested the method on a diverse set of open shell molecules and condensed phase systems. The molecular test set includes \(\mathrm{O}_2\) (triplet), \(\mathrm{NO}\) (doublet), \(\mathrm{CuCl}_2\) (doublet), and the benzene dimer cation \((\mathrm{C}_6\mathrm{H}_6)_2^{+}\) (doublet). The condensed phase test set comprises a high-spin polaron configuration on the rutile \(\mathrm{TiO}_2\)(110) surface with one oxygen vacancy (multiplicity \(=3\)), the rutile \(\mathrm{TiO}_2\)(110) surface with one adsorbed OH group (multiplicity \(=2\)), antiferromagnetic \(\mathrm{NiO}\) and \(\mathrm{CoO}\) (multiplicity \(=1\)), a high-spin polaron configuration of bulk \(\mathrm{CeO}_2\) containing one oxygen vacancy (OV) (multiplicity \(=3\)), and a snapshot from an AIMD simulation containing 32 liquid water molecules and one \(\mathrm{Ag}^{2+}\) ion (multiplicity \(=2\)).
\subsection{4.1 Reproduction Accuracy of Target Electron Densities}
\begin{figure}[H]
\centering

\subfigure[]{
\includegraphics[width=\textwidth]{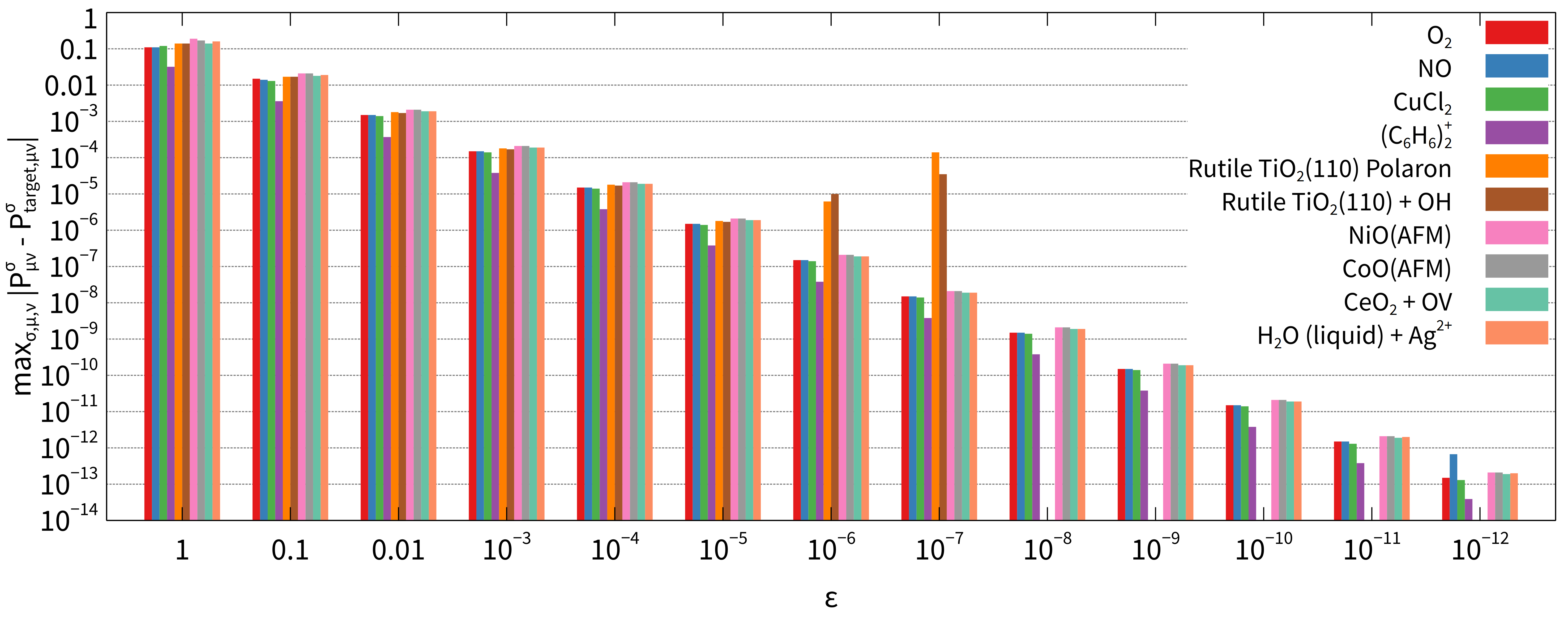}
\label{fig2a}
}
\subfigure[]{
\includegraphics[width=\textwidth]{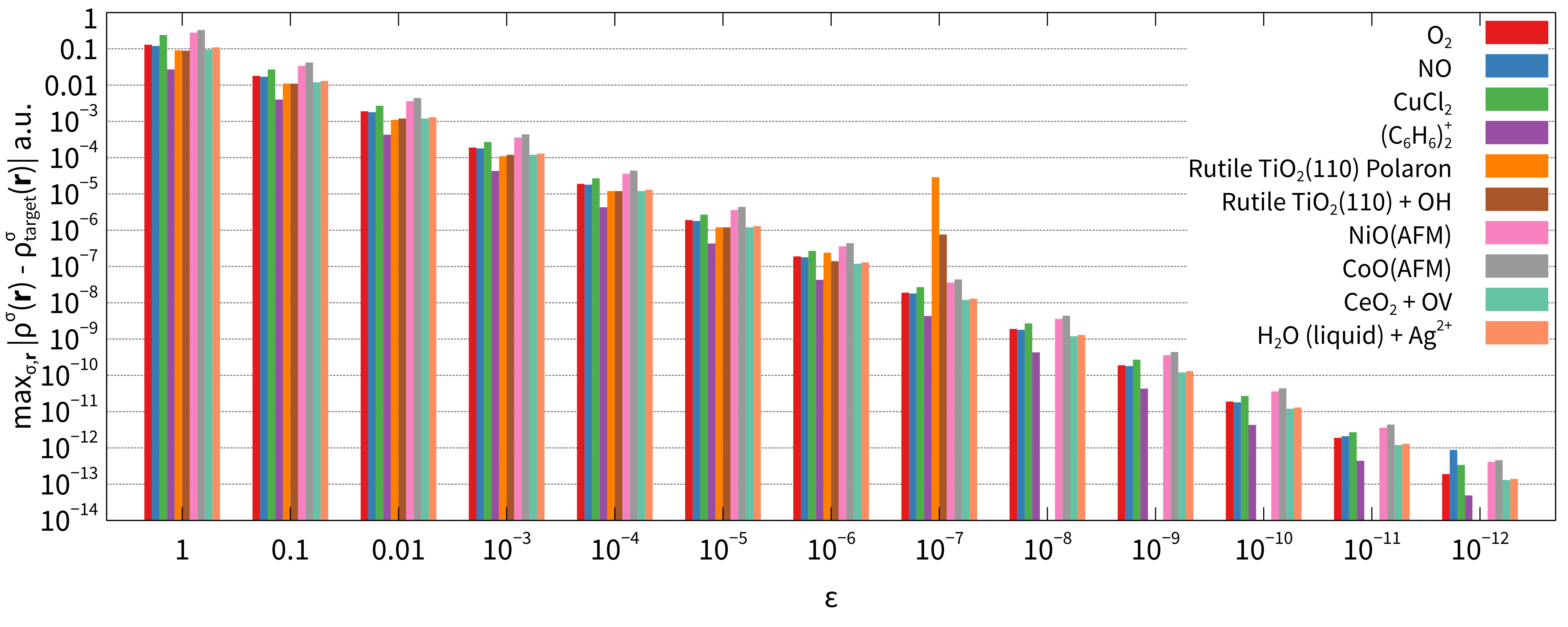}
\label{fig2b}
}

\caption{(a) The maximum absolute deviations among all matrix elements of \(\Delta \mathbf{P}'^{\uparrow}\) and \(\Delta \mathbf{P}'^{\downarrow}\), and (b) the maximum absolute deviations over all real space grid points of \(\rho^{\uparrow}(\mathbf{r})-\rho_{\mathrm{target}}^{\uparrow}(\mathbf{r})\) and \(\rho^{\downarrow}(\mathbf{r})-\rho_{\mathrm{target}}^{\downarrow}(\mathbf{r})\), for each test system at a given value of \(\epsilon\). When \(\epsilon\) is smaller than \(1\times10^{-7}\), omitted bars indicate calculations that did not finish within the wall-time limit. We note that, for each test system, once \(\epsilon\) becomes smaller than the value listed below, the SCF convergence criterion can no longer be satisfied within 6000 SCF iterations. The values shown in the figure therefore correspond to those obtained at the 6000th SCF iteration: \(\mathrm{O}_2\) (\(\epsilon=1\times10^{-10}\)), \(\mathrm{NO}\) (\(\epsilon=1\times10^{-10}\)), \(\mathrm{CuCl}_2\) (\(\epsilon=1\times10^{-11}\)), \((\mathrm{C}_6\mathrm{H}_6)_2^{+}\) (\(\epsilon=1\times10^{-11}\)), rutile \(\mathrm{TiO}_2(110)\) polaron (\(\epsilon=1\times10^{-4}\)), rutile \(\mathrm{TiO}_2(110)+\mathrm{OH}\) (\(\epsilon=1\times10^{-3}\)), \(\mathrm{NiO}\) (AFM) (\(\epsilon=1\times10^{-11}\)), \(\mathrm{CoO}\) (AFM) (\(\epsilon=1\times10^{-11}\)), bulk \(\mathrm{CeO}_2\) polaron (\(\epsilon=1\times10^{-11}\)), and \(32\,\mathrm{H_2O}+\mathrm{Ag}^{2+}\) (\(\epsilon=1\times10^{-11}\)).}\label{fig2}
\end{figure}

Ensuring high accuracy in reproducing both the electron density matrix and the corresponding real space electron density is essential. To assess the performance of the method, we report the residual errors in the resulting density matrix and real space electron density with respect to their target counterparts. For each test system, a forward PBE KS-DFT calculation was first performed to obtain the target electron density matrix. Subsequently, a sequence of inverse KS-DFT SCF optimizations with progressively decreasing \(\epsilon\) (Fig.~\ref{fig1}) was carried out using the OT-DIIS optimizer. In each inverse KS-DFT SCF optimization, the parameter \texttt{ds\_min}, which controls the optimization step size, was fixed at \(0.01/\epsilon\) to ensure stable convergence.\\
Fig.~\ref{fig2a} shows the maximum absolute deviations over all elements of \(\Delta \mathbf{P}'^{\uparrow}\) and \(\Delta \mathbf{P}'^{\downarrow}\), while Fig.~\ref{fig2b} shows the maximum absolute deviations over all real space grid points of \(\rho^{\uparrow}(\mathbf{r})-\rho_{\mathrm{target}}^{\uparrow}(\mathbf{r})\) and \(\rho^{\downarrow}(\mathbf{r})-\rho_{\mathrm{target}}^{\downarrow}(\mathbf{r})\) for each test system at each value of \(\epsilon\). The numerical data shown in the figures are listed in Tables S1 and S2 in Section~2 of the Supporting Information, respectively. Here \(\rho^{\sigma}(\mathbf{r})\) denotes the real space electron density of spin channel \(\sigma=\uparrow\) or \(\sigma=\downarrow\) at the end of the SCF optimization, and \(\rho_{\mathrm{target}}^{\sigma}(\mathbf{r})\) denotes the corresponding target density. The two figures show that, as \(\epsilon\) decreases, the maximum deviations in both the density matrix and the real space electron density decrease approximately linearly on a logarithmic scale. Except for the two \(\mathrm{TiO}_2\) surface systems, which exhibit convergence difficulties, reducing \(\epsilon\) to the order of \(10^{-12}\) lowers the deviations to the order of \(10^{-12}\)–\(10^{-13}\) for the density matrix and \(10^{-12}\)–\(10^{-13}\,\mathrm{a.u.}\) for the real space electron density.\\
\subsection{4.2 Comparison with the ZMP Method in terms of the Attainable Limit of Accuracy}
\begin{figure}[H]
\centering

\subfigure[]{
\includegraphics[width=\textwidth]{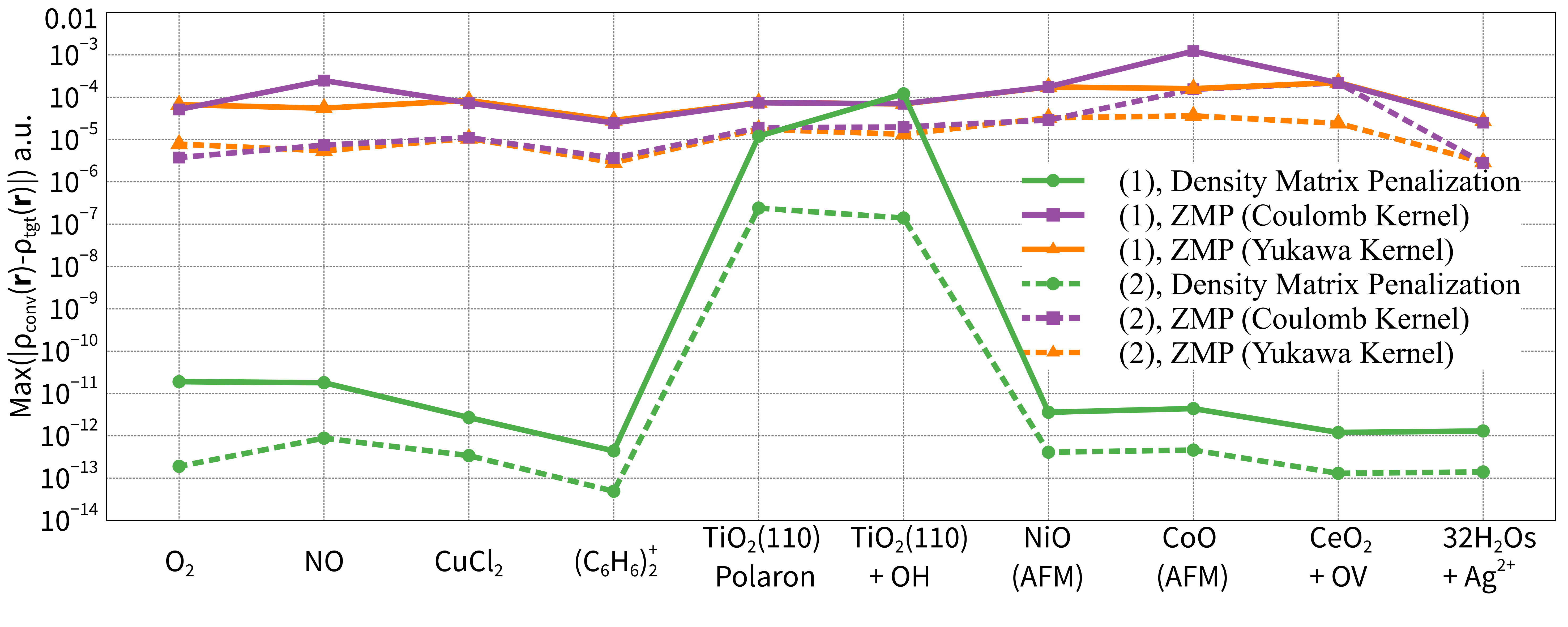}
\label{fig3a}
}
\subfigure[]{
\includegraphics[width=\textwidth]{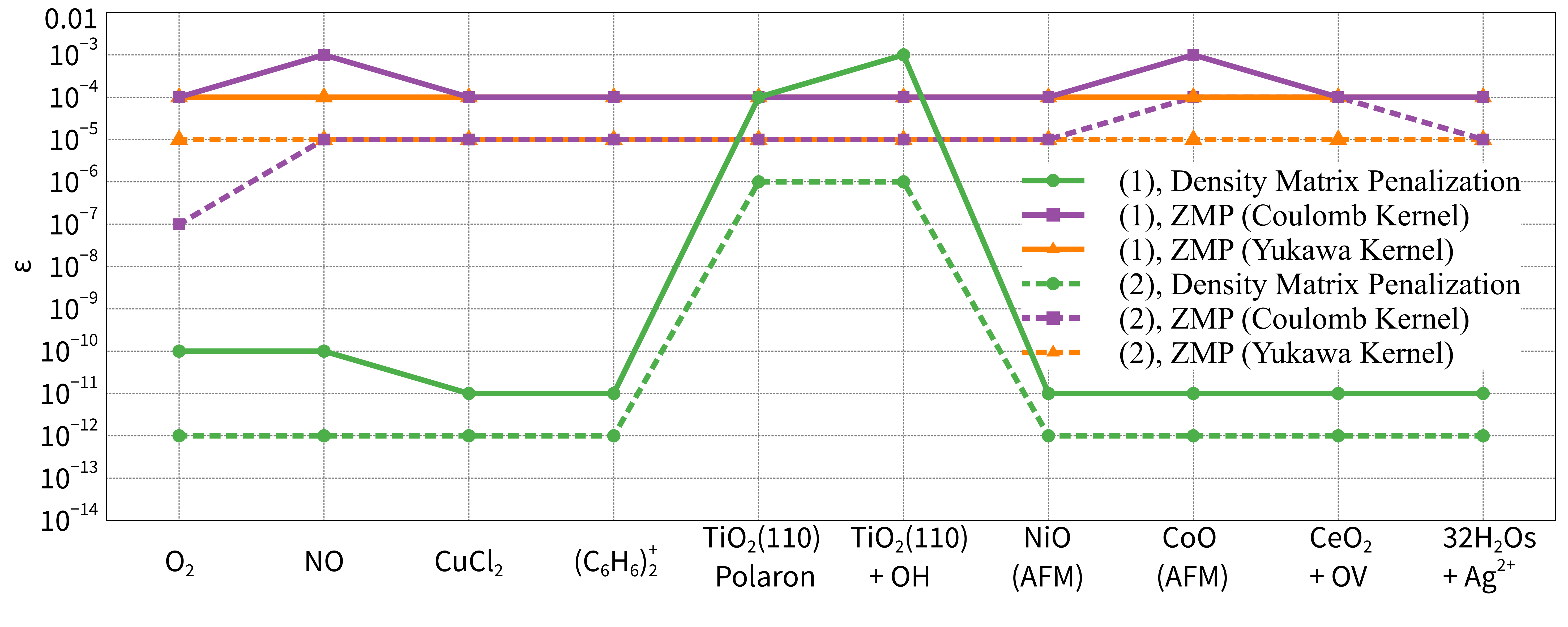}
\label{fig3b}
}

\caption{(a) Smallest attainable maximum absolute deviation between the real space electron density and the target density on the grid for all test systems as \(\epsilon\) decreases. (b) Corresponding \(\epsilon\) values. Results obtained using the density matrix penalization method, the Coulomb-based ZMP method, and the Yukawa-based ZMP method are shown in green, blue, and yellow, respectively. Solid and dashed lines denote calculations that converged within 6000 SCF iterations and calculations that reached the maximum limit of 6000 iterations without satisfying the convergence criterion, respectively. Detailed numerical results for the Coulomb-based and Yukawa-based ZMP methods are listed in Tables S3 and S4.}\label{fig3}
\end{figure}

As \(\epsilon\) approaches zero, the converged electron density should, in principle, approach the target density. In practical inverse KS-DFT calculations, however, a limit to the attainable accuracy is often observed. To assess this limit, we compared the proposed approach with the ZMP method. For this purpose, we implemented a spin-unrestricted version of the ZMP method in CP2K, with \(\epsilon\) updated according to the scheme described in Section~2.4 and illustrated in Fig.~\ref{fig1}. The only difference lies in how the penalty potential matrix is constructed before solving the KS equations. In the ZMP implementation, the penalty potential is first constructed in real space from the density difference using an FFT-based evaluation of the Hartree (or Yukawa) potential (see Section~1 of the Supporting Information)\cite{ZMP_pra_1994, Herbst_prb_2025}. The resulting potential is then integrated with the Gaussian basis functions, and the corresponding matrix elements are added to the Hamiltonian matrix.\\
Fig.~\ref{fig3a} shows, for all test systems, the smallest attainable maximum absolute deviation between the real space electron density and the target density on the grid as \(\epsilon\) decreases, while Fig.~\ref{fig3b} shows the corresponding \(\epsilon\) values. Results obtained using the density matrix penalization method, the Coulomb-based ZMP method, and the Yukawa-based ZMP method are indicated by different colors. Solid and dashed lines denote results for calculations that converged within 6000 SCF iterations and calculations that reached the maximum limit of 6000 iterations without satisfying the convergence criterion, respectively. As shown in Fig.~\ref{fig3a}, for all systems except the two \(\mathrm{TiO}_2\) surface systems, the density matrix penalization method achieves the smallest attainable maximum absolute deviations, which are \(6\text{-}8\) orders of magnitude lower than those obtained with the Coulomb-based and Yukawa-based ZMP methods. Another indication of this significant improvement in accuracy is the smallest value of \(\epsilon\) for which SCF convergence can still be achieved. As shown in Fig.~\ref{fig3b}, for these systems, the smallest attainable \(\epsilon\) obtained with the density matrix penalization method is \(6\text{-}8\) orders of magnitude smaller than that obtained with the ZMP method. For the two \(\mathrm{TiO}_2\) surface systems, although none of the methods can achieve SCF convergence within 6000 iterations when \(\epsilon < 10^{-4}\), further decreasing \(\epsilon\) still allows the density matrix penalization method to achieve the smallest attainable maximum absolute deviations, about two orders of magnitude lower than those obtained with the ZMP methods. A slight improvement in the attainable accuracy of the ZMP method is also observed in Fig.~\ref{fig3a} and Fig.~\ref{fig3b} when the Coulomb kernel is replaced by the Yukawa kernel. Additional tests further show that the grid resolution has little impact on the attainable accuracy of the ZMP method in our calculations (see Section 5 of the Supporting Information).
\subsection{4.3 Computational Efficiency and Convergence}
\begin{figure}[H]
\centering
\includegraphics[width=\textwidth]{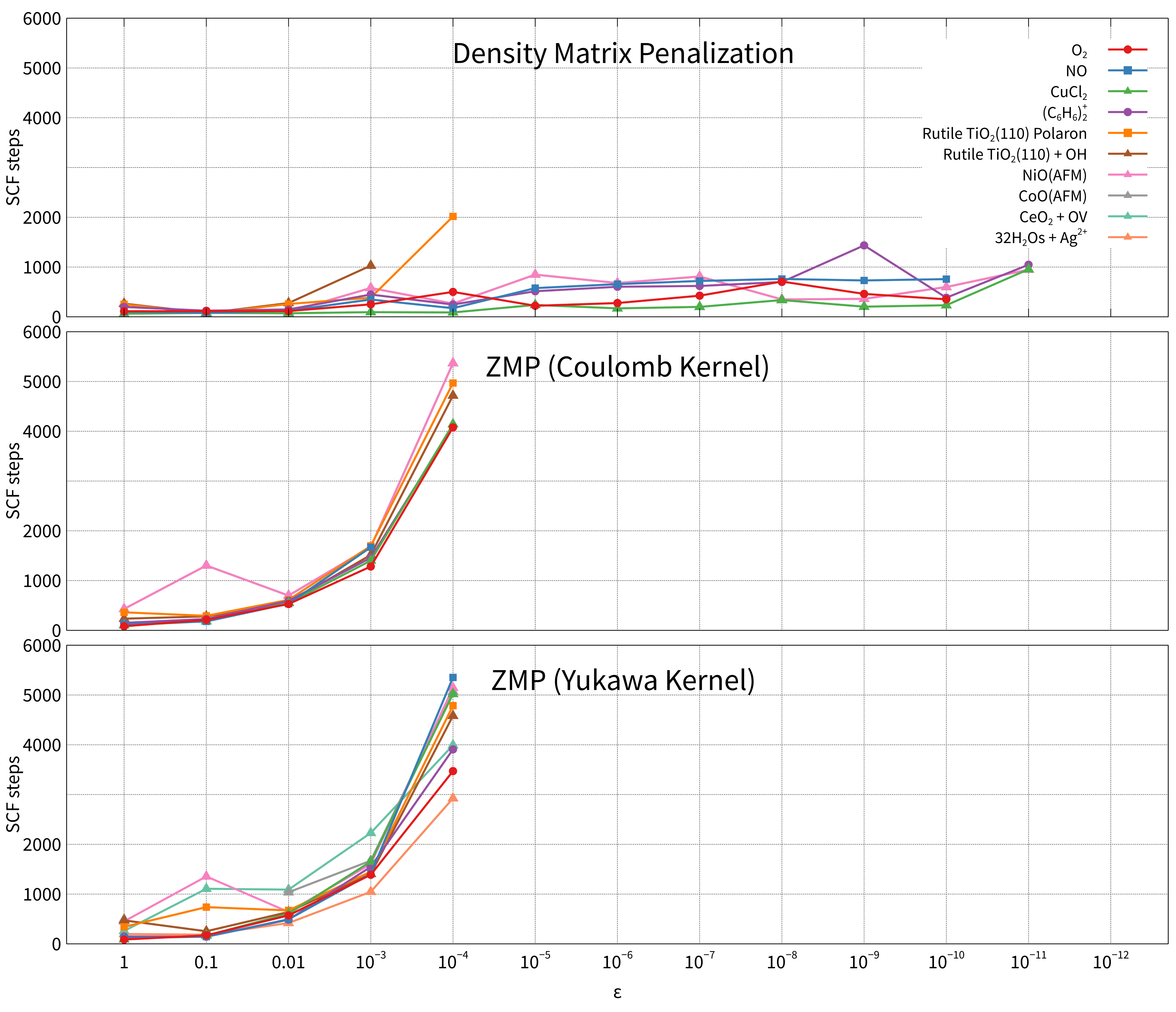}
\caption{Number of SCF iterations required to achieve SCF convergence for all test systems in inverse KS-DFT calculations along the decreasing sequence of \(\epsilon\). Results obtained using the density matrix penalization method, the Coulomb-based ZMP method, and the Yukawa-based ZMP method are shown in the top, middle, and bottom panels, respectively.}\label{fig4}
\end{figure}

Figure~\ref{fig4} reports, for all test systems, the numbers of SCF iterations required to achieve SCF convergence using the density matrix penalization method, the Coulomb-based ZMP method, and the Yukawa-based ZMP method along the decreasing sequence of \(\epsilon\) in the inverse KS-DFT calculations. One can observe that, for the ZMP methods, the number of iterations required for SCF convergence increases rapidly and becomes noticeably larger than that required for the  density matrix penalization method when \(\epsilon\) is reduced below \(0.01\). As shown in the figure, over the range of \(\epsilon\) from \(1\) to \(1\times10^{-11}\), most calculations using the density matrix penalization method achieve SCF convergence within 1000 iterations. For the ZMP schemes, no calculations achieve SCF convergence once \(\epsilon\) becomes smaller than \(1\times10^{-4}\). At \(\epsilon = 1\times10^{-12}\), none of the calculations converge, suggesting that the numerical noise floor of the present setup has been reached. The convergence status and the corresponding numbers of SCF iterations for all calculations are listed in Tables S5-7. We also note that additional tests show that the grid resolution has little impact on the convergence behavior of the ZMP calculations or on the number of SCF iterations required for convergence (see Section 5 of the Supporting Information).\\
\section{5. Conclusions}
We have developed a density matrix based inverse KS-DFT scheme that is formulated entirely within a finite Gaussian basis representation. By defining the density matrix mismatch in the Löwdin orthogonalized basis, we obtain a penalty energy that is invariant under unitary rotations in that representation. The corresponding penalty potential contribution to the KS Hamiltonian is then derived analytically in the original Gaussian basis. The inversion is driven by a ZMP/MY-type penalty tightening strategy through decreasing \(\epsilon\), while maintaining a representation consistent optimization in the Gaussian basis algebra.\\
The method has been implemented in CP2K for spin-unrestricted calculations and benchmarked on a diverse set of open shell systems, including molecules, condensed phase models, and surfaces. For most systems, progressively tightening the penalty drives the electron density into accurate agreement with the target, reaching \(\max_{\sigma,\mathbf{r}}|\rho^{\sigma}(\mathbf{r})-\rho_{\mathrm{target}}^{\sigma}(\mathbf{r})|\sim 1\times10^{-12}\text{-}1\times10^{-13}\) a.u. Compared with a conventional ZMP implementation in Gaussian basis KS-DFT based on real space density differences, the attainable density matching accuracy is improved by \(6\text{-}8\) orders of magnitude, and the SCF convergence remains robust over a much wider range of penalty strengths. In particular, most calculations converge within 1000 SCF iterations for \(\epsilon\) decreased from \(1\) down to \(1\times10^{-10}\) (or \(1\times10^{-11}\)), whereas the ZMP approach fails to reach SCF convergence once \(\epsilon<10^{-4}\) in the present setup, with rapidly increasing iteration counts as \(\epsilon\) decreases. For the two rutile \(\mathrm{TiO}_2\)(110) surface test cases, although all methods encounter convergence difficulties at small \(\epsilon\), the present approach still achieves noticeably smaller attainable density deviations than the ZMP approach.\\
Overall, the method substantially improves attainable density matching accuracy over conventional ZMP and remains convergent over a much wider range of penalty strengths for most systems tested here. A systematic analysis of the reconstructed potentials, their dependence on regularization choices and basis set completeness, and improved convergence for difficult surface cases will be pursued in future work.\\

\begin{acknowledgement}

This work was supported by the University of Zurich and SNSF Spark Project 228719. This work was supported by the grants from the Swiss National Supercomputing Centre (CSCS) under project ID lp89 and lp11.\\

\end{acknowledgement}


\bibliography{achemso-demo}
\end{document}


\renewcommand{\thepage}{S\arabic{page}}
\begin{suppinfo}
\setcounter{equation}{0}
\renewcommand\theequation{S\arabic{equation}}
\setcounter{figure}{0}
\renewcommand\thefigure{S\arabic{figure}}
\setcounter{table}{0}
\renewcommand\thetable{S\arabic{table}}
\subsection{1. Implementation of the ZMP method}
This part presents the formulas used in our implementation of ZMP in CP2K.
Penalty energy:
\begin{equation}
E_P=\sum_\sigma \frac{1}{\epsilon} \iint \frac{\left(\rho^\sigma(\boldsymbol{r})-\rho_{\text {target }}^\sigma(\boldsymbol{r})\right)\left(\rho^\sigma\left(\boldsymbol{r}^{\prime}\right)-\rho_{\text {target }}^\sigma\left(\boldsymbol{r}^{\prime}\right)\right)}{\left|\boldsymbol{r}-\boldsymbol{r}^{\prime}\right|} d \boldsymbol{r} d \boldsymbol{r}^{\prime}.
\label{eqs1}
\end{equation}
Penalty potential (Coulomb flavor) in real space for spin channel $\sigma$:
\begin{equation}
V_P^\sigma(\boldsymbol{r})=\frac{2}{\epsilon} \int \frac{\rho^\sigma\left(\boldsymbol{r}^{\prime}\right)-\rho_{\text {target }}^\sigma\left(\boldsymbol{r}^{\prime}\right)}{\left|\boldsymbol{r}-\boldsymbol{r}^{\prime}\right|} d \boldsymbol{r}^{\prime},
\label{eqs2}
\end{equation}
and penalty potential (Yukawa flavor) in real space for spin channel $\sigma$:
\begin{equation}
V_P^\sigma(\boldsymbol{r})=\frac{2}{\epsilon} \int \frac{\rho^\sigma\left(\boldsymbol{r}^{\prime}\right)-\rho_{\text {target }}^\sigma\left(\boldsymbol{r}^{\prime}\right)}{\left|\boldsymbol{r}-\boldsymbol{r}^{\prime}\right|} e^{-\left|\boldsymbol{r}-\boldsymbol{r}^{\prime}\right|} d \boldsymbol{r}^{\prime} .
\label{eqs3}
\end{equation}
The above real space penalty potentials can be calculated efficiently in CP2K via FFT. Penalty potential matrix for spin channel $\sigma$:
\begin{equation}
\left(K_P^\sigma\right)_{a b}=\int \phi_a(\boldsymbol{r}) V_P^\sigma(\boldsymbol{r}) \phi_b(\boldsymbol{r}) d \boldsymbol{r}.
\label{eqs4}
\end{equation}
\subsection{2. The data shown in Figs.2a-2b}
Red numbers indicate values obtained at the 6000th SCF iteration, where the SCF procedure did not converge to the convergence criterion (\texttt{EPS\_SCF} = \(1\times10^{-6}\)) within 6000 iterations. The symbol \(\times\) indicates calculations that did not finish within the time limit.\\
\begin{table}[H]
\centering
\caption{The data shown in Fig.2a: maximum absolute density matrix deviation as a function of \(\epsilon\).}
\label{tab:dm_dev}
\small
\begin{adjustbox}{width=\textwidth}
\begin{tabular}{lcccccccccc}
\toprule
\multirow{2}{*}{\(\epsilon\)} &
\multicolumn{10}{c}{\(\displaystyle \max_{\sigma,\mu,\nu}\left|P_{\mu\nu}^{\sigma}-P_{\mathrm{target},\mu\nu}^{\sigma}\right|\)} \\
\cmidrule(lr){2-11}
& \(\mathrm{O}_2\) & \(\mathrm{NO}\) & \(\mathrm{CuCl}_2\) & \((\mathrm{C}_6\mathrm{H}_6)_2^{+}\) & \(\mathrm{TiO}_2\) polaron & \(\mathrm{TiO}_2+\mathrm{OH}\) & \(\mathrm{NiO}\) & \(\mathrm{CoO}\) & \(\mathrm{CeO}_2+\mathrm{OV}\) & \(32\,\mathrm{H_2O}+\mathrm{Ag}^{2+}\) \\
\midrule
\(1.0\times10^{0}\)   & \(1.1\times10^{-1}\) & \(1.1\times10^{-1}\) & \(1.2\times10^{-1}\) & \(3.2\times10^{-2}\) & \(1.4\times10^{-1}\) & \(1.4\times10^{-1}\) & \(1.9\times10^{-1}\) & \(1.7\times10^{-1}\) & \(1.4\times10^{-1}\) & \(1.6\times10^{-1}\) \\
\(1.0\times10^{-1}\)  & \(1.5\times10^{-2}\) & \(1.4\times10^{-2}\) & \(1.3\times10^{-2}\) & \(3.6\times10^{-3}\) & \(1.7\times10^{-2}\) & \(1.7\times10^{-2}\) & \(2.1\times10^{-2}\) & \(2.1\times10^{-2}\) & \(1.8\times10^{-2}\) & \(1.9\times10^{-2}\) \\
\(1.0\times10^{-2}\)  & \(1.5\times10^{-3}\) & \(1.5\times10^{-3}\) & \(1.4\times10^{-3}\) & \(3.7\times10^{-4}\) & \(1.8\times10^{-3}\) & \(1.7\times10^{-3}\) & \(2.1\times10^{-3}\) & \(2.1\times10^{-3}\) & \(1.9\times10^{-3}\) & \(1.9\times10^{-3}\) \\
\(1.0\times10^{-3}\)  & \(1.5\times10^{-4}\) & \(1.5\times10^{-4}\) & \(1.4\times10^{-4}\) & \(3.8\times10^{-5}\) & \(1.8\times10^{-4}\) & \(1.7\times10^{-4}\) & \(2.1\times10^{-4}\) & \(2.1\times10^{-4}\) & \(1.9\times10^{-4}\) & \(1.9\times10^{-4}\) \\
\(1.0\times10^{-4}\)  & \(1.5\times10^{-5}\) & \(1.5\times10^{-5}\) & \(1.4\times10^{-5}\) & \(3.8\times10^{-6}\) & \(1.8\times10^{-5}\) & \textcolor{red}{\(1.7\times10^{-5}\)}& \(2.1\times10^{-5}\) & \(2.1\times10^{-5}\) & \(1.9\times10^{-5}\) & \(1.9\times10^{-5}\) \\
\(1.0\times10^{-5}\)  & \(1.5\times10^{-6}\) & \(1.5\times10^{-6}\) & \(1.4\times10^{-6}\) & \(3.8\times10^{-7}\) & \textcolor{red}{\(1.8\times10^{-6}\)}& \textcolor{red}{\(1.7\times10^{-6}\)}& \(2.1\times10^{-6}\) & \(2.1\times10^{-6}\) & \(1.9\times10^{-6}\) & \(1.9\times10^{-6}\) \\
\(1.0\times10^{-6}\)  & \(1.5\times10^{-7}\) & \(1.5\times10^{-7}\) & \(1.4\times10^{-7}\) & \(3.8\times10^{-8}\) & \textcolor{red}{\(6.2\times10^{-6}\)}& \textcolor{red}{\(1.0\times10^{-5}\)}& \(2.1\times10^{-7}\) & \(2.1\times10^{-7}\) & \(1.9\times10^{-7}\) & \(1.9\times10^{-7}\) \\
\(1.0\times10^{-7}\)  & \(1.5\times10^{-8}\) & \(1.5\times10^{-8}\) & \(1.4\times10^{-8}\) & \(3.8\times10^{-9}\) & \textcolor{red}{\(1.4\times10^{-4}\)}& \textcolor{red}{\(3.5\times10^{-5}\)}& \(2.1\times10^{-8}\) & \(2.1\times10^{-8}\) & \(1.9\times10^{-8}\) & \(1.9\times10^{-8}\) \\
\(1.0\times10^{-8}\)  & \(1.5\times10^{-9}\) & \(1.5\times10^{-9}\) & \(1.4\times10^{-9}\) & \(3.8\times10^{-10}\) & \(\times\) & \(\times\) & \(2.1\times10^{-9}\) & \(2.1\times10^{-9}\) & \(1.9\times10^{-9}\) & \(1.9\times10^{-9}\) \\
\(1.0\times10^{-9}\)  & \(1.5\times10^{-10}\) & \(1.5\times10^{-10}\) & \(1.4\times10^{-10}\) & \(3.8\times10^{-11}\) & \(\times\) & \(\times\) & \(2.1\times10^{-10}\) & \(2.1\times10^{-10}\) & \(1.9\times10^{-10}\) & \(1.9\times10^{-10}\) \\
\(1.0\times10^{-10}\) & \(1.5\times10^{-11}\) & \(1.5\times10^{-11}\) & \(1.4\times10^{-11}\) & \(3.8\times10^{-12}\) & \(\times\) & \(\times\) & \(2.1\times10^{-11}\) & \(2.1\times10^{-11}\) & \(1.9\times10^{-11}\) & \(1.9\times10^{-11}\) \\
\(1.0\times10^{-11}\) & \textcolor{red}{\(1.5\times10^{-12}\)}& \textcolor{red}{\(1.5\times10^{-12}\)}& \(1.3\times10^{-12}\) & \(3.8\times10^{-13}\) & \(\times\) & \(\times\) & \(2.1\times10^{-12}\) & \(2.1\times10^{-12}\) & \(1.9\times10^{-12}\) & \(2.0\times10^{-12}\) \\
\(1.0\times10^{-12}\) & \textcolor{red}{\(1.5\times10^{-13}\)}& \textcolor{red}{\(6.7\times10^{-13}\)}& \textcolor{red}{\(1.3\times10^{-13}\)}& \textcolor{red}{\(3.9\times10^{-14}\)}& \(\times\) & \(\times\) & \textcolor{red}{\(2.1\times10^{-13}\)}& \textcolor{red}{\(2.1\times10^{-13}\)}& \textcolor{red}{\(1.9\times10^{-13}\)}& \textcolor{red}{\(2.0\times10^{-13}\)}\\
\bottomrule
\end{tabular}
\end{adjustbox}
\end{table}

\begin{table}[H]
\centering
\caption{The data shown in Fig.2b: maximum absolute real space electron density deviation as a function of \(\epsilon\).}
\label{tab:rho_dev}
\small
\begin{adjustbox}{width=\textwidth}
\begin{tabular}{lcccccccccc}
\toprule
\multirow{2}{*}{\(\epsilon\)} &
\multicolumn{10}{c}{\(\displaystyle \max_{\sigma,\mathbf{r}}\left|\rho^{\sigma}(\mathbf{r})-\rho_{\mathrm{target}}^{\sigma}(\mathbf{r})\right|\)} \\
\cmidrule(lr){2-11}
& \(\mathrm{O}_2\) & \(\mathrm{NO}\) & \(\mathrm{CuCl}_2\) & \((\mathrm{C}_6\mathrm{H}_6)_2^{+}\) & \(\mathrm{TiO}_2\) polaron & \(\mathrm{TiO}_2+\mathrm{OH}\) & \(\mathrm{NiO}\) & \(\mathrm{CoO}\) & \(\mathrm{CeO}_2+\mathrm{OV}\) & \(32\,\mathrm{H_2O}+\mathrm{Ag}^{2+}\) \\
\midrule
\(1.0\times10^{0}\)   & \(1.3\times10^{-1}\) & \(1.2\times10^{-1}\) & \(2.4\times10^{-1}\) & \(2.7\times10^{-2}\) & \(9.1\times10^{-2}\) & \(8.9\times10^{-2}\) & \(2.8\times10^{-1}\) & \(3.3\times10^{-1}\) & \(9.5\times10^{-2}\) & \(1.1\times10^{-1}\) \\
\(1.0\times10^{-1}\)  & \(1.8\times10^{-2}\) & \(1.7\times10^{-2}\) & \(2.7\times10^{-2}\) & \(4.0\times10^{-3}\) & \(1.1\times10^{-2}\) & \(1.1\times10^{-2}\) & \(3.4\times10^{-2}\) & \(4.2\times10^{-2}\) & \(1.2\times10^{-2}\) & \(1.3\times10^{-2}\) \\
\(1.0\times10^{-2}\)  & \(1.9\times10^{-3}\) & \(1.8\times10^{-3}\) & \(2.7\times10^{-3}\) & \(4.3\times10^{-4}\) & \(1.1\times10^{-3}\) & \(1.2\times10^{-3}\) & \(3.6\times10^{-3}\) & \(4.4\times10^{-3}\) & \(1.2\times10^{-3}\) & \(1.3\times10^{-3}\) \\
\(1.0\times10^{-3}\)  & \(1.9\times10^{-4}\) & \(1.8\times10^{-4}\) & \(2.7\times10^{-4}\) & \(4.3\times10^{-5}\) & \(1.1\times10^{-4}\) & \(1.2\times10^{-4}\) & \(3.6\times10^{-4}\) & \(4.4\times10^{-4}\) & \(1.2\times10^{-4}\) & \(1.3\times10^{-4}\) \\
\(1.0\times10^{-4}\)  & \(1.9\times10^{-5}\) & \(1.8\times10^{-5}\) & \(2.7\times10^{-5}\) & \(4.3\times10^{-6}\) & \(1.2\times10^{-5}\) & \textcolor{red}{\(1.2\times10^{-5}\)}& \(3.6\times10^{-5}\) & \(4.4\times10^{-5}\) & \(1.2\times10^{-5}\) & \(1.3\times10^{-5}\) \\
\(1.0\times10^{-5}\)  & \(1.9\times10^{-6}\) & \(1.8\times10^{-6}\) & \(2.7\times10^{-6}\) & \(4.3\times10^{-7}\) & \textcolor{red}{\(1.2\times10^{-6}\)}& \textcolor{red}{\(1.2\times10^{-6}\)}& \(3.6\times10^{-6}\) & \(4.4\times10^{-6}\) & \(1.2\times10^{-6}\) & \(1.3\times10^{-6}\) \\
\(1.0\times10^{-6}\)  & \(1.9\times10^{-7}\) & \(1.8\times10^{-7}\) & \(2.7\times10^{-7}\) & \(4.3\times10^{-8}\) & \textcolor{red}{\(2.4\times10^{-7}\)}& \textcolor{red}{\(1.4\times10^{-7}\)}& \(3.6\times10^{-7}\) & \(4.4\times10^{-7}\) & \(1.2\times10^{-7}\) & \(1.3\times10^{-7}\) \\
\(1.0\times10^{-7}\)  & \(1.9\times10^{-8}\) & \(1.8\times10^{-8}\) & \(2.7\times10^{-8}\) & \(4.3\times10^{-9}\) & \textcolor{red}{\(2.9\times10^{-5}\)}& \textcolor{red}{\(7.6\times10^{-7}\)}& \(3.6\times10^{-8}\) & \(4.4\times10^{-8}\) & \(1.2\times10^{-8}\) & \(1.3\times10^{-8}\) \\
\(1.0\times10^{-8}\)  & \(1.9\times10^{-9}\) & \(1.8\times10^{-9}\) & \(2.7\times10^{-9}\) & \(4.3\times10^{-10}\) & \(\times\) & \(\times\) & \(3.6\times10^{-9}\) & \(4.4\times10^{-9}\) & \(1.2\times10^{-9}\) & \(1.3\times10^{-9}\) \\
\(1.0\times10^{-9}\)  & \(1.9\times10^{-10}\) & \(1.8\times10^{-10}\) & \(2.7\times10^{-10}\) & \(4.3\times10^{-11}\) & \(\times\) & \(\times\) & \(3.6\times10^{-10}\) & \(4.4\times10^{-10}\) & \(1.2\times10^{-10}\) & \(1.3\times10^{-10}\) \\
\(1.0\times10^{-10}\) & \(1.9\times10^{-11}\) & \(1.8\times10^{-11}\) & \(2.7\times10^{-11}\) & \(4.3\times10^{-12}\) & \(\times\) & \(\times\) & \(3.6\times10^{-11}\) & \(4.4\times10^{-11}\) & \(1.2\times10^{-11}\) & \(1.3\times10^{-11}\) \\
\(1.0\times10^{-11}\) & \textcolor{red}{\(1.9\times10^{-12}\)}& \textcolor{red}{\(2.1\times10^{-12}\)}& \(2.7\times10^{-12}\) & \(4.4\times10^{-13}\) & \(\times\) & \(\times\) & \(3.6\times10^{-12}\) & \(4.4\times10^{-12}\) & \(1.2\times10^{-12}\) & \(1.3\times10^{-12}\) \\
\(1.0\times10^{-12}\) & \textcolor{red}{\(1.9\times10^{-13}\)}& \textcolor{red}{\(8.8\times10^{-13}\)}& \textcolor{red}{\(3.4\times10^{-13}\)}& \textcolor{red}{\(4.9\times10^{-14}\)}& \(\times\) & \(\times\) & \textcolor{red}{\(4.1\times10^{-13}\)}& \textcolor{red}{\(4.6\times10^{-13}\)}& \textcolor{red}{\(1.3\times10^{-13}\)}& \textcolor{red}{\(1.4\times10^{-13}\)}\\
\bottomrule
\end{tabular}
\end{adjustbox}
\end{table}
\subsection{3. \(\max_{\sigma,\mu,\nu}\left|P_{\mu\nu}^{\sigma}-P_{\mathrm{target},\mu\nu}^{\sigma}\right|\) and \(\max_{\sigma,\mathbf r}\left|\rho^{\sigma}(\mathbf r)-\rho_{\mathrm{target}}^{\sigma}(\mathbf r)\right|\)
results calculated using the ZMP schemes}

Red numbers indicate values obtained at the 6000th SCF iteration, where the SCF procedure did not converge to the convergence criterion (\texttt{EPS\_SCF} = \(1\times10^{-6}\)) within 6000 iterations. The symbol \(\times\) indicates calculations that did not finish within the time limit.\\
\begin{table}[H]
\centering
\caption{The results calculated by using Coulomb-based ZMP method: maximum absolute real space electron density deviation as a function of \(\epsilon\).}
\label{stab3}
\small
\begin{adjustbox}{width=\textwidth}
\begin{tabular}{lcccccccccc}
\toprule
\multirow{2}{*}{\(\epsilon\)} &
\multicolumn{10}{c}{\(\displaystyle \max_{\sigma,\mathbf r}\left|\rho^{\sigma}(\mathbf r)-\rho_{\mathrm{target}}^{\sigma}(\mathbf r)\right|\)} \\
\cmidrule(lr){2-11}
& \(\mathrm{O}_2\) & \(\mathrm{NO}\) & \(\mathrm{CuCl}_2\) & \((\mathrm{C}_6\mathrm{H}_6)_2^{+}\) & \(\mathrm{TiO}_2\) polaron & \(\mathrm{TiO}_2+\mathrm{OH}\) & \(\mathrm{NiO}\) & \(\mathrm{CoO}\) & \(\mathrm{CeO}_2+\mathrm{OV}\) & \(32\,\mathrm{H_2O}+\mathrm{Ag}^{2+}\) \\
\midrule
\(1.0\times10^{0}\)   & \(1.4\times10^{-1}\) & \(1.5\times10^{-1}\) & \(3.3\times10^{-1}\) & \(4.4\times10^{-2}\) & \(2.2\times10^{-1}\) & \(1.3\times10^{-1}\) & \(4.2\times10^{-1}\) & \(5.9\times10^{-1}\) & \(2.9\times10^{-1}\) & \(1.6\times10^{-1}\) \\
\(1.0\times10^{-1}\)  & \(2.0\times10^{-2}\) & \(2.2\times10^{-2}\) & \(4.3\times10^{-2}\) & \(1.3\times10^{-2}\) & \(3.4\times10^{-2}\) & \(1.8\times10^{-2}\) & \(7.8\times10^{-2}\) & \(8.4\times10^{-2}\) & \(6.7\times10^{-2}\) & \(2.3\times10^{-2}\) \\
\(1.0\times10^{-2}\)  & \(1.9\times10^{-3}\) & \(2.2\times10^{-3}\) & \(6.2\times10^{-3}\) & \(1.7\times10^{-3}\) & \(3.8\times10^{-3}\) & \(2.0\times10^{-3}\) & \(1.0\times10^{-2}\) & \(1.1\times10^{-2}\) & \(1.3\times10^{-2}\) & \(2.0\times10^{-3}\) \\
\(1.0\times10^{-3}\)  & \(3.6\times10^{-4}\) & \(2.5\times10^{-4}\) & \(6.9\times10^{-4}\) & \(2.1\times10^{-4}\) & \(5.1\times10^{-4}\) & \(4.2\times10^{-4}\) & \(1.1\times10^{-3}\) & \(1.2\times10^{-3}\) & \(1.9\times10^{-3}\) & \(2.4\times10^{-4}\) \\
\(1.0\times10^{-4}\)  & \(5.1\times10^{-5}\) & \textcolor{red}{\(3.1\times10^{-5}\)}& \(7.3\times10^{-5}\) & \(2.5\times10^{-5}\) & \(7.4\times10^{-5}\) & \(7.0\times10^{-5}\) & \(1.8\times10^{-4}\) & \textcolor{red}{\(1.5\times10^{-4}\)}& \(2.2\times10^{-4}\) & \(2.5\times10^{-5}\) \\
\(1.0\times10^{-5}\)  & \textcolor{red}{\(8.7\times10^{-6}\)} & \textcolor{red}{\(7.4\times10^{-6}\)} & \textcolor{red}{\(1.1\times10^{-5}\)} & \textcolor{red}{\(3.6\times10^{-6}\)} & \textcolor{red}{\(1.9\times10^{-5}\)} & \textcolor{red}{\(2.0\times10^{-5}\)} & \textcolor{red}{\(2.9\times10^{-5}\)} & \(\times\) & \(\times\) & \textcolor{red}{\(2.8\times10^{-6}\)} \\
\(1.0\times10^{-6}\)  & \textcolor{red}{\(9.5\times10^{-6}\)} & \textcolor{red}{\(1.4\times10^{-5}\)} & \textcolor{red}{\(1.1\times10^{-5}\)} & \textcolor{red}{\(6.0\times10^{-6}\)} & \(\times\) & \(\times\) & \(\times\) & \(\times\) & \(\times\) & \textcolor{red}{\(5.6\times10^{-6}\)} \\
\(1.0\times10^{-7}\)  & \textcolor{red}{\(3.8\times10^{-6}\)} & \textcolor{red}{\(5.0\times10^{-5}\)} & \textcolor{red}{\(1.1\times10^{-4}\)} & \textcolor{red}{\(7.5\times10^{-6}\)} & \(\times\) & \(\times\) & \(\times\) & \(\times\) & \(\times\) & \textcolor{red}{\(1.8\times10^{-5}\)} \\
\(1.0\times10^{-8}\)  & \textcolor{red}{\(1.0\times10^{-3}\)} & \textcolor{red}{\(2.7\times10^{-5}\)} & \textcolor{red}{\(9.4\times10^{-5}\)} & \textcolor{red}{\(7.0\times10^{-5}\)} & \(\times\) & \(\times\) & \(\times\) & \(\times\) & \(\times\) & \textcolor{red}{\(2.6\times10^{-5}\)} \\
\(1.0\times10^{-9}\)  & \textcolor{red}{\(1.0\times10^{-3}\)} & \textcolor{red}{\(5.9\times10^{-5}\)} & \textcolor{red}{\(6.7\times10^{-5}\)} & \textcolor{red}{\(1.8\times10^{-3}\)} & \(\times\) & \(\times\) & \(\times\) & \(\times\) & \(\times\) & \textcolor{red}{\(4.4\times10^{-5}\)} \\
\(1.0\times10^{-10}\) & \textcolor{red}{\(3.4\times10^{-2}\)} & \textcolor{red}{\(7.1\times10^{-2}\)} & \textcolor{red}{\(2.2\times10^{-4}\)} & \textcolor{red}{\(1.8\times10^{-3}\)} & \(\times\) & \(\times\) & \(\times\) & \(\times\) & \(\times\) & \textcolor{red}{\(2.5\times10^{-5}\)} \\
\(1.0\times10^{-11}\) & \textcolor{red}{\(6.3\times10^{-1}\)} & \textcolor{red}{\(6.0\times10^{-1}\)} & \textcolor{red}{\(5.0\times10^{-1}\)} & \textcolor{red}{\(1.5\times10^{-1}\)} & \(\times\) & \(\times\) & \(\times\) & \(\times\) & \(\times\) & \textcolor{red}{\(2.4\times10^{-5}\)} \\
\(1.0\times10^{-12}\) & \textcolor{red}{\(1.1\times10^{0}\)} & \textcolor{red}{\(3.4\times10^{-1}\)} & \textcolor{red}{\(9.4\times10^{-1}\)} & \textcolor{red}{\(4.0\times10^{-1}\)} & \(\times\) & \(\times\) & \(\times\) & \(\times\) & \(\times\) & \textcolor{red}{\(2.6\times10^{-1}\)} \\
\bottomrule
\end{tabular}
\end{adjustbox}
\end{table}
\begin{table}[H]
\centering
\caption{The results calculated by using Yukawa-based ZMP method: maximum absolute real space electron density deviation as a function of \(\epsilon\).}
\label{stab4}
\small
\begin{adjustbox}{width=\textwidth}
\begin{tabular}{lcccccccccc}
\toprule
\multirow{2}{*}{\(\epsilon\)} &
\multicolumn{10}{c}{\(\displaystyle \max_{\sigma,\mathbf r}\left|\rho^{\sigma}(\mathbf r)-\rho_{\mathrm{target}}^{\sigma}(\mathbf r)\right|\)} \\
\cmidrule(lr){2-11}
& \(\mathrm{O}_2\) & \(\mathrm{NO}\) & \(\mathrm{CuCl}_2\) & \((\mathrm{C}_6\mathrm{H}_6)_2^{+}\) & \(\mathrm{TiO}_2\) polaron & \(\mathrm{TiO}_2+\mathrm{OH}\) & \(\mathrm{NiO}\) & \(\mathrm{CoO}\) & \(\mathrm{CeO}_2+\mathrm{OV}\) & \(32\,\mathrm{H_2O}+\mathrm{Ag}^{2+}\) \\
\midrule
\(1.0\times10^{0}\)   & \(1.8\times10^{-1}\) & \(1.9\times10^{-1}\) & \(5.0\times10^{-1}\) & \(4.8\times10^{-2}\) & \(3.4\times10^{-1}\) & \(1.6\times10^{-1}\) & \(4.8\times10^{-1}\) & \(7.0\times10^{-1}\) & \(4.0\times10^{-1}\) & \(2.0\times10^{-1}\) \\
\(1.0\times10^{-1}\)  & \(2.5\times10^{-2}\) & \(2.7\times10^{-2}\) & \(5.8\times10^{-2}\) & \(1.6\times10^{-2}\) & \(4.2\times10^{-2}\) & \(2.7\times10^{-2}\) & \(8.7\times10^{-2}\) & \textcolor{red}{\(9.6\times10^{-2}\)}& \(7.3\times10^{-2}\) & \(3.3\times10^{-2}\) \\
\(1.0\times10^{-2}\)  & \(2.2\times10^{-3}\) & \(2.5\times10^{-3}\) & \(7.0\times10^{-3}\) & \(1.5\times10^{-3}\) & \(4.3\times10^{-3}\) & \(3.1\times10^{-3}\) & \(1.1\times10^{-2}\) & \(1.2\times10^{-2}\) & \(1.3\times10^{-2}\) & \(2.8\times10^{-3}\) \\
\(1.0\times10^{-3}\)  & \(3.8\times10^{-4}\) & \(3.0\times10^{-4}\) & \(7.5\times10^{-4}\) & \(2.1\times10^{-4}\) & \(5.5\times10^{-4}\) & \(4.2\times10^{-4}\) & \(1.2\times10^{-3}\) & \(1.3\times10^{-3}\) & \(1.9\times10^{-3}\) & \(2.6\times10^{-4}\) \\
\(1.0\times10^{-4}\)  & \(6.7\times10^{-5}\) & \(5.5\times10^{-5}\) & \(8.4\times10^{-5}\) & \(2.8\times10^{-5}\) & \(7.5\times10^{-5}\) & \(6.9\times10^{-5}\) & \(1.7\times10^{-4}\) & \(1.6\times10^{-4}\) & \(2.2\times10^{-4}\) & \(2.7\times10^{-5}\) \\
\(1.0\times10^{-5}\)  & \textcolor{red}{\(7.8\times10^{-6}\)} & \textcolor{red}{\(5.4\times10^{-6}\)} & \textcolor{red}{\(1.1\times10^{-5}\)} & \textcolor{red}{\(2.9\times10^{-6}\)} & \textcolor{red}{\(1.8\times10^{-5}\)} & \textcolor{red}{\(1.3\times10^{-5}\)} & \textcolor{red}{\(3.3\times10^{-5}\)} & \textcolor{red}{\(3.6\times10^{-5}\)} & \textcolor{red}{\(2.4\times10^{-5}\)} & \textcolor{red}{\(2.9\times10^{-6}\)} \\
\(1.0\times10^{-6}\)  & \textcolor{red}{\(8.1\times10^{-6}\)} & \textcolor{red}{\(9.5\times10^{-5}\)} & \textcolor{red}{\(3.4\times10^{-5}\)} & \textcolor{red}{\(3.8\times10^{-6}\)} & \(\times\) & \(\times\) & \(\times\) & \(\times\) & \(\times\) & \textcolor{red}{\(3.1\times10^{-6}\)} \\
\(1.0\times10^{-7}\)  & \textcolor{red}{\(1.1\times10^{-5}\)} & \textcolor{red}{\(5.9\times10^{-5}\)} & \textcolor{red}{\(7.0\times10^{-5}\)} & \textcolor{red}{\(7.0\times10^{-6}\)} & \(\times\) & \(\times\) & \(\times\) & \(\times\) & \(\times\) & \textcolor{red}{\(7.0\times10^{-6}\)} \\
\(1.0\times10^{-8}\)  & \textcolor{red}{\(3.0\times10^{-5}\)} & \textcolor{red}{\(5.6\times10^{-5}\)} & \textcolor{red}{\(7.0\times10^{-5}\)} & \textcolor{red}{\(2.9\times10^{-5}\)} & \(\times\) & \(\times\) & \(\times\) & \(\times\) & \(\times\) & \textcolor{red}{\(1.3\times10^{-5}\)} \\
\(1.0\times10^{-9}\)  & \textcolor{red}{\(1.6\times10^{-5}\)} & \textcolor{red}{\(7.0\times10^{-5}\)} & \textcolor{red}{\(7.2\times10^{-5}\)} & \textcolor{red}{\(7.4\times10^{-4}\)} & \(\times\) & \(\times\) & \(\times\) & \(\times\) & \(\times\) & \textcolor{red}{\(1.3\times10^{-3}\)} \\
\(1.0\times10^{-10}\) & \textcolor{red}{\(1.2\times10^{-4}\)} & \textcolor{red}{\(1.2\times10^{-4}\)} & \textcolor{red}{\(3.4\times10^{-4}\)} & \textcolor{red}{\(2.2\times10^{-1}\)} & \(\times\) & \(\times\) & \(\times\) & \(\times\) & \(\times\) & \textcolor{red}{\(1.1\times10^{-3}\)} \\
\(1.0\times10^{-11}\) & \textcolor{red}{\(8.2\times10^{-4}\)} & \textcolor{red}{\(4.4\times10^{-2}\)} & \textcolor{red}{\(9.1\times10^{-4}\)} & \textcolor{red}{\(4.8\times10^{-1}\)} & \(\times\) & \(\times\) & \(\times\) & \(\times\) & \(\times\) & \textcolor{red}{\(8.0\times10^{-4}\)} \\
\(1.0\times10^{-12}\) & \textcolor{red}{\(5.0\times10^{-1}\)} & \textcolor{red}{\(1.4\times10^{0}\)} & \textcolor{red}{\(6.7\times10^{-2}\)} & \textcolor{red}{\(3.9\times10^{-1}\)} & \(\times\) & \(\times\) & \(\times\) & \(\times\) & \(\times\) & \textcolor{red}{\(6.7\times10^{-1}\)} \\
\bottomrule
\end{tabular}
\end{adjustbox}
\end{table}

\subsection{4. The number of SCF iterations required to achieve SCF convergence shown in Fig. 4}

The symbol “\(\times\)” indicates calculations for which the SCF iterations did not satisfy the convergence criterion (\texttt{EPS\_SCF} = \(1\times10^{-6}\)) within 6000 iterations or did not finish within the time limit.

\begin{table}[H]
\centering
\caption{The results for the density matrix penalization method: number of SCF iterations as a function of \(\epsilon\).}
\label{stab5}
\small
\begin{adjustbox}{width=\textwidth}
\begin{tabular}{lcccccccccc}
\toprule
\multirow{2}{*}{\(\epsilon\)} &
\multicolumn{10}{c}{SCF steps} \\
\cmidrule(lr){2-11}
& \(\mathrm{O}_2\) & \(\mathrm{NO}\) & \(\mathrm{CuCl}_2\) & \((\mathrm{C}_6\mathrm{H}_6)_2^{+}\) & \(\mathrm{TiO}_2\) polaron & \(\mathrm{TiO}_2+\mathrm{OH}\) & \(\mathrm{NiO}\) & \(\mathrm{CoO}\) & \(\mathrm{CeO}_2+\mathrm{OV}\) & \(32\,\mathrm{H_2O}+\mathrm{Ag}^{2+}\) \\
\midrule
\(1.0\times10^{0}\)   & 112 & 104 & 63  & 199  & 252  & 266  & 260  & 232  & 267  & 176 \\
\(1.0\times10^{-1}\)  & 114 & 77  & 80  & 123  & 67   & 73   & 84   & 74   & 183  & 57  \\
\(1.0\times10^{-2}\)  & 116 & 123 & 72  & 149  & 258  & 277  & 96   & 303  & 219  & 240 \\
\(1.0\times10^{-3}\)  & 251 & 348 & 94  & 448  & 385  & 1029 & 580  & 178  & 326  & 100 \\
\(1.0\times10^{-4}\)  & 501 & 173 & 88  & 247  & 2017 & \(\times\) & 265  & 129  & 421  & 151 \\
\(1.0\times10^{-5}\)  & 222 & 577 & 239 & 514  & \(\times\) & \(\times\) & 848  & 291  & 375  & 451 \\
\(1.0\times10^{-6}\)  & 277 & 657 & 170 & 604  & \(\times\) & \(\times\) & 679  & 253  & 203  & 217 \\
\(1.0\times10^{-7}\)  & 425 & 721 & 200 & 620  & \(\times\) & \(\times\) & 811  & 252  & 376  & 238 \\
\(1.0\times10^{-8}\)  & 709 & 761 & 337 & 700  & \(\times\) & \(\times\) & 350  & 268  & 701  & 270 \\
\(1.0\times10^{-9}\)  & 461 & 731 & 204 & 1435 & \(\times\) & \(\times\) & 361  & 351  & 887  & 296 \\
\(1.0\times10^{-10}\) & 352 & 758 & 233 & 384  & \(\times\) & \(\times\) & 599  & 345  & 620  & 306 \\
\(1.0\times10^{-11}\) & \(\times\) & \(\times\) & 956 & 1047 & \(\times\) & \(\times\) & 948  & 347  & 1193 & 358 \\
\(1.0\times10^{-12}\) & \(\times\) & \(\times\) & \(\times\) & \(\times\) & \(\times\) & \(\times\) & \(\times\) & \(\times\) & \(\times\) & \(\times\) \\
\bottomrule
\end{tabular}
\end{adjustbox}
\end{table}

\begin{table}[H]
\centering
\caption{The results for the Coulomb-based ZMP method: number of SCF iterations as a function of \(\epsilon\).}
\label{stab6}
\small
\begin{adjustbox}{width=\textwidth}
\begin{tabular}{lcccccccccc}
\toprule
\multirow{2}{*}{\(\epsilon\)} &
\multicolumn{10}{c}{SCF steps} \\
\cmidrule(lr){2-11}
& \(\mathrm{O}_2\) & \(\mathrm{NO}\) & \(\mathrm{CuCl}_2\) & \((\mathrm{C}_6\mathrm{H}_6)_2^{+}\) & \(\mathrm{TiO}_2\) polaron & \(\mathrm{TiO}_2+\mathrm{OH}\) & \(\mathrm{NiO}\) & \(\mathrm{CoO}\) & \(\mathrm{CeO}_2+\mathrm{OV}\) & \(32\,\mathrm{H_2O}+\mathrm{Ag}^{2+}\) \\
\midrule
\(1.0\times10^{0}\)   & 82   & 120  & 102  & 150  & 362  & 234  & 430  & 522  & 322  & 196 \\
\(1.0\times10^{-1}\)  & 211  & 179  & 182  & 228  & 292  & 281  & 1303 & 4987 & 5171 & 203 \\
\(1.0\times10^{-2}\)  & 529  & 535  & 549  & 594  & 612  & 559  & 699  & 862  & 1248 & 470 \\
\(1.0\times10^{-3}\)  & 1283 & 1669 & 1400 & 1457 & 1696 & 1515 & 1676 & 2050 & 2181 & 1165 \\
\(1.0\times10^{-4}\)  & 4078 & \(\times\) & 4142 & 4100 & 4969 & 4717 & 5369 & \(\times\) & 4977 & 3252 \\
\(1.0\times10^{-5}\)  & \(\times\) & \(\times\) & \(\times\) & \(\times\) & \(\times\) & \(\times\) & \(\times\) & \(\times\) & \(\times\) & \(\times\) \\
\(1.0\times10^{-6}\)  & \(\times\) & \(\times\) & \(\times\) & \(\times\) & \(\times\) & \(\times\) & \(\times\) & \(\times\) & \(\times\) & \(\times\) \\
\(1.0\times10^{-7}\)  & \(\times\) & \(\times\) & \(\times\) & \(\times\) & \(\times\) & \(\times\) & \(\times\) & \(\times\) & \(\times\) & \(\times\) \\
\(1.0\times10^{-8}\)  & \(\times\) & \(\times\) & \(\times\) & \(\times\) & \(\times\) & \(\times\) & \(\times\) & \(\times\) & \(\times\) & \(\times\) \\
\(1.0\times10^{-9}\)  & \(\times\) & \(\times\) & \(\times\) & \(\times\) & \(\times\) & \(\times\) & \(\times\) & \(\times\) & \(\times\) & \(\times\) \\
\(1.0\times10^{-10}\) & \(\times\) & \(\times\) & \(\times\) & \(\times\) & \(\times\) & \(\times\) & \(\times\) & \(\times\) & \(\times\) & \(\times\) \\
\(1.0\times10^{-11}\) & \(\times\) & \(\times\) & \(\times\) & \(\times\) & \(\times\) & \(\times\) & \(\times\) & \(\times\) & \(\times\) & \(\times\) \\
\(1.0\times10^{-12}\) & \(\times\) & \(\times\) & \(\times\) & \(\times\) & \(\times\) & \(\times\) & \(\times\) & \(\times\) & \(\times\) & \(\times\) \\
\bottomrule
\end{tabular}
\end{adjustbox}
\end{table}

\begin{table}[H]
\centering
\caption{The results for the Yukawa-based ZMP method: number of SCF iterations as a function of \(\epsilon\).}
\label{stab7}
\small
\begin{adjustbox}{width=\textwidth}
\begin{tabular}{lcccccccccc}
\toprule
\multirow{2}{*}{\(\epsilon\)} &
\multicolumn{10}{c}{SCF steps} \\
\cmidrule(lr){2-11}
& \(\mathrm{O}_2\) & \(\mathrm{NO}\) & \(\mathrm{CuCl}_2\) & \((\mathrm{C}_6\mathrm{H}_6)_2^{+}\) & \(\mathrm{TiO}_2\) polaron & \(\mathrm{TiO}_2+\mathrm{OH}\) & \(\mathrm{NiO}\) & \(\mathrm{CoO}\) & \(\mathrm{CeO}_2+\mathrm{OV}\) & \(32\,\mathrm{H_2O}+\mathrm{Ag}^{2+}\) \\
\midrule
\(1.0\times10^{0}\)   & 87   & 117  & 91   & 149  & 341  & 472  & 452  & 493  & 260  & 198 \\
\(1.0\times10^{-1}\)  & 171  & 145  & 167  & 143  & 738  & 253  & 1355 & \(\times\) & 1107 & 184 \\
\(1.0\times10^{-2}\)  & 572  & 493  & 605  & 486  & 672  & 641  & 641  & 1037 & 1090 & 419 \\
\(1.0\times10^{-3}\)  & 1388 & 1407 & 1652 & 1548 & 1451 & 1405 & 1603 & 1672 & 2229 & 1046 \\
\(1.0\times10^{-4}\)  & 3471 & 5353 & 5021 & 3908 & 4786 & 4582 & 5145 & 5070 & 3992 & 2924 \\
\(1.0\times10^{-5}\)  & \(\times\) & \(\times\) & \(\times\) & \(\times\) & \(\times\) & \(\times\) & \(\times\) & \(\times\) & \(\times\) & \(\times\) \\
\(1.0\times10^{-6}\)  & \(\times\) & \(\times\) & \(\times\) & \(\times\) & \(\times\) & \(\times\) & \(\times\) & \(\times\) & \(\times\) & \(\times\) \\
\(1.0\times10^{-7}\)  & \(\times\) & \(\times\) & \(\times\) & \(\times\) & \(\times\) & \(\times\) & \(\times\) & \(\times\) & \(\times\) & \(\times\) \\
\(1.0\times10^{-8}\)  & \(\times\) & \(\times\) & \(\times\) & \(\times\) & \(\times\) & \(\times\) & \(\times\) & \(\times\) & \(\times\) & \(\times\) \\
\(1.0\times10^{-9}\)  & \(\times\) & \(\times\) & \(\times\) & \(\times\) & \(\times\) & \(\times\) & \(\times\) & \(\times\) & \(\times\) & \(\times\) \\
\(1.0\times10^{-10}\) & \(\times\) & \(\times\) & \(\times\) & \(\times\) & \(\times\) & \(\times\) & \(\times\) & \(\times\) & \(\times\) & \(\times\) \\
\(1.0\times10^{-11}\) & \(\times\) & \(\times\) & \(\times\) & \(\times\) & \(\times\) & \(\times\) & \(\times\) & \(\times\) & \(\times\) & \(\times\) \\
\(1.0\times10^{-12}\) & \(\times\) & \(\times\) & \(\times\) & \(\times\) & \(\times\) & \(\times\) & \(\times\) & \(\times\) & \(\times\) & \(\times\) \\
\bottomrule
\end{tabular}
\end{adjustbox}
\end{table}

\subsection{5. Test of different CUTOFF and REL\_CUTOFF for the Yukawa-based ZMP method}
To assess the potential impact of real space grid resolution on the electrostatic potential evaluation, and consequently on the accuracy and convergence behavior of ZMP inversion, inverse DFT calculations were performed for the NO molecule using the Yukawa-based ZMP method. The resulting density deviations, SCF convergence behavior, and the numbers of SCF iterations are summarized in Tables~S8 and~S9. It can be seen that increasing the \texttt{CUTOFF} and \texttt{REL\_CUTOFF} parameters leads to almost no noticeable change in either the density deviation or the number of SCF iterations. These results indicate that, within the tested range, the grid resolution does not constitute a direct bottleneck for the accuracy or convergence of the ZMP method.

\begin{table}[H]
\centering
\caption{\(\max_{\sigma,\mathbf r}\left|\rho^{\sigma}(\mathbf r)-\rho_{\mathrm{target}}^{\sigma}(\mathbf r)\right|\) 
(in a.u.) calculated using the Yukawa-based ZMP scheme for the NO molecule with increasing 
\texttt{CUTOFF} and \texttt{REL\_CUTOFF} values (in Ry). 
Red numbers indicate values obtained at the 6000th SCF iteration, where the SCF procedure did not converge 
to the convergence criterion (\texttt{EPS\_SCF} = \(1\times10^{-6}\)) within 6000 iterations. 
The symbol "\(\times\)'' indicates calculations that did not finish within the time limit.}
\label{stab8}
\small
\begin{adjustbox}{width=0.75\textwidth}
\begin{tabular}{lccccc}
\toprule
\(\epsilon\) &
\multicolumn{5}{c}{\(\max_{\sigma,\mathbf r}\left|\rho^{\sigma}(\mathbf r)-\rho_{\mathrm{target}}^{\sigma}(\mathbf r)\right|\)} \\
\cmidrule(lr){2-6}
 & (320, 40) & (500, 50) & (600, 60) & (700, 70) & (800, 80) \\
\midrule
\(1.0\times10^{0}\)  & \(1.9\times10^{-1}\) & \(1.9\times10^{-1}\) & \(1.9\times10^{-1}\) & \(1.9\times10^{-1}\) & \(1.9\times10^{-1}\) \\
\(1.0\times10^{-1}\) & \(2.7\times10^{-2}\) & \(2.7\times10^{-2}\) & \(2.7\times10^{-2}\) & \(2.8\times10^{-2}\) & \(2.8\times10^{-2}\) \\
\(1.0\times10^{-2}\) & \(2.5\times10^{-3}\) & \(2.5\times10^{-3}\) & \(2.5\times10^{-3}\) & \(2.6\times10^{-3}\) & \(2.5\times10^{-3}\) \\
\(1.0\times10^{-3}\) & \(3.0\times10^{-4}\) & \(2.9\times10^{-4}\) & \(3.0\times10^{-4}\) & \(3.0\times10^{-4}\) & \(3.0\times10^{-4}\) \\
\(1.0\times10^{-4}\) & \(5.5\times10^{-5}\) & \(5.2\times10^{-5}\) & \(5.0\times10^{-5}\) & \(5.4\times10^{-5}\) & \(5.1\times10^{-5}\) \\
\(1.0\times10^{-5}\) & \textcolor{red}{\(5.4\times10^{-6}\)} & \textcolor{red}{\(5.6\times10^{-6}\)} & \textcolor{red}{\(6.1\times10^{-6}\)} & \textcolor{red}{\(5.2\times10^{-6}\)} & \textcolor{red}{\(6.0\times10^{-6}\)} \\
\(1.0\times10^{-6}\) & \textcolor{red}{\(9.5\times10^{-5}\)} & \textcolor{red}{\(6.1\times10^{-5}\)} & \textcolor{red}{\(4.5\times10^{-5}\)} & \textcolor{red}{\(9.3\times10^{-5}\)} & \textcolor{red}{\(2.0\times10^{-5}\)} \\
\(1.0\times10^{-7}\) & \textcolor{red}{\(5.9\times10^{-5}\)} & \textcolor{red}{\(2.6\times10^{-5}\)} & \textcolor{red}{\(1.3\times10^{-5}\)} & \textcolor{red}{\(4.4\times10^{-5}\)} & \textcolor{red}{\(1.2\times10^{-5}\)} \\
\(1.0\times10^{-8}\) & \textcolor{red}{\(5.6\times10^{-5}\)} & \textcolor{red}{\(2.1\times10^{-5}\)} & \textcolor{red}{\(2.8\times10^{-4}\)} & \textcolor{red}{\(3.7\times10^{-5}\)} & \textcolor{red}{\(4.1\times10^{-3}\)} \\
\(1.0\times10^{-9}\) & \textcolor{red}{\(7.0\times10^{-5}\)} & \textcolor{red}{\(4.3\times10^{-5}\)} & \textcolor{red}{\(3.1\times10^{-4}\)} & \textcolor{red}{\(3.7\times10^{-5}\)} & \textcolor{red}{\(4.5\times10^{-3}\)} \\
\(1.0\times10^{-10}\)& \textcolor{red}{\(1.2\times10^{-4}\)} & \textcolor{red}{\(1.1\times10^{-4}\)} & \textcolor{red}{\(4.5\times10^{-4}\)} & \textcolor{red}{\(1.1\times10^{-4}\)} & \textcolor{red}{\(4.1\times10^{-3}\)} \\
\(1.0\times10^{-11}\)& \textcolor{red}{\(4.4\times10^{-2}\)} & \textcolor{red}{\(5.3\times10^{-4}\)} & \textcolor{red}{\(1.9\times10^{-3}\)} & \textcolor{red}{\(6.7\times10^{-1}\)} & \textcolor{red}{\(5.5\times10^{-3}\)} \\
\(1.0\times10^{-12}\)& \textcolor{red}{\(1.4\times10^{0}\)} & \textcolor{red}{\(1.7\times10^{-1}\)} & \textcolor{red}{\(5.4\times10^{-2}\)} & \textcolor{red}{\(8.6\times10^{-1}\)} & \textcolor{red}{\(1.4\times10^{0}\)} \\
\bottomrule
\end{tabular}
\end{adjustbox}
\end{table}

\begin{table}[H]
\centering
\caption{The number of SCF iterations required to achieve SCF convergence for the NO molecule with increasing \texttt{CUTOFF} and \texttt{REL\_CUTOFF} values (in Ry). The symbol "\(\times\)'' indicates calculations for which the SCF procedure did not converge to the convergence criterion (\texttt{EPS\_SCF} = \(1\times10^{-6}\)) within 6000 iterations.}
\label{stab9}
\small
\begin{adjustbox}{width=0.7\textwidth}
\begin{tabular}{lccccc}
\toprule
\multirow{2}{*}{\(\epsilon\)} &
\multicolumn{5}{c}{\texttt{CUTOFF}, \texttt{REL\_CUTOFF} (Ry)} \\
\cmidrule(lr){2-6}
 & (320, 40) & (500, 50) & (600, 60) & (700, 70) & (800, 80) \\
\midrule
\(1.0\times10^{0}\)  & 117  & 124  & 124  & 114  & 113 \\
\(1.0\times10^{-1}\) & 145  & 143  & 143  & 152  & 186 \\
\(1.0\times10^{-2}\) & 493  & 468  & 468  & 491  & 474 \\
\(1.0\times10^{-3}\) & 1407 & 1393 & 1393 & 1413 & 1439 \\
\(1.0\times10^{-4}\) & 5353 & 4553 & 4553 & 4379 & 5254 \\
\(1.0\times10^{-5}\) & \(\times\) & \(\times\) & \(\times\) & \(\times\) & \(\times\) \\
\(1.0\times10^{-6}\) & \(\times\) & \(\times\) & \(\times\) & \(\times\) & \(\times\) \\
\(1.0\times10^{-7}\) & \(\times\) & \(\times\) & \(\times\) & \(\times\) & \(\times\) \\
\(1.0\times10^{-8}\) & \(\times\) & \(\times\) & \(\times\) & \(\times\) & \(\times\) \\
\(1.0\times10^{-9}\) & \(\times\) & \(\times\) & \(\times\) & \(\times\) & \(\times\) \\
\(1.0\times10^{-10}\)& \(\times\) & \(\times\) & \(\times\) & \(\times\) & \(\times\) \\
\(1.0\times10^{-11}\)& \(\times\) & \(\times\) & \(\times\) & \(\times\) & \(\times\) \\
\(1.0\times10^{-12}\)& \(\times\) & \(\times\) & \(\times\) & \(\times\) & \(\times\) \\
\bottomrule
\end{tabular}
\end{adjustbox}
\end{table}
\subsection{6. Structures}
Coordinates are given in \AA\ and angles are given in degrees ($^\circ$). 'ABC' and 'ALPHA BETA GAMMA' obey the convention of CP2K input. The following structure files are in 'xyz' format.\\

\textbf{$\mathrm{O}_2$ (triplet)}
\begin{verbatim}
2
O2 (triplet) r(O=O)=1.208 Å, centered on origin, along z. 
O  0.000000  0.000000 -0.604000
O  0.000000  0.000000  0.604000
\end{verbatim}
-----------------------------------------------\\
Periodic Box: ABC 12 12 12; ALPHA BETA GAMMA 90 90 90\\

\textbf{$\mathrm{NO}$ (doublet)}
\begin{verbatim}
2
NO (doublet) r(N=O)=1.154 Å, centered on origin, along z
N  0.000000  0.000000 -0.577000
O  0.000000  0.000000  0.577000
\end{verbatim}
-----------------------------------------------\\
Periodic Box: ABC 12 12 12; ALPHA BETA GAMMA 90 90 90\\

\textbf{$\mathrm{CuCl}_2$ (doublet)}
\begin{verbatim}
3
CuCl2 (doublet) linear guess, r(Cu-Cl)=2.07 Å, along z
Cl  0.000000  0.000000 -2.070000
Cu  0.000000  0.000000  0.000000
Cl  0.000000  0.000000  2.070000
\end{verbatim}
-----------------------------------------------\\
Periodic Box: ABC 12 12 12; ALPHA BETA GAMMA 90 90 90\\

\textbf{The benzene dimer cation $(\mathrm{C}_6\mathrm{H}_6)_2^{+}$ (doublet)}
\begin{verbatim}
24
(C6H6)2+
C  0.004479  1.590102  1.393307
C -0.004479 -1.590102  1.393307
C  0.004479  1.590102 -1.393307
C -0.004479 -1.590102 -1.393307
C  1.161486  2.025138  0.692992
C -1.161486 -2.025138  0.692992
C  1.161486  2.025138 -0.692992
C -1.161486 -2.025138 -0.692992
C -1.161486  1.198031  0.695030
C  1.161486 -1.198031  0.695030
C -1.161486  1.198031 -0.695030
C  1.161486 -1.198031 -0.695030
H  0.000428  1.603775  2.485837
H -0.000428 -1.603775  2.485837
H  0.000428  1.603775 -2.485837
H -0.000428 -1.603775 -2.485837
H  2.038171  2.358855  1.250683
H -2.038171 -2.358855  1.250683
H  2.038171  2.358855 -1.250683
H -2.038171 -2.358855 -1.250683
H -2.056555  0.917010  1.251775
H  2.056555 -0.917010  1.251775
H -2.056555  0.917010 -1.251775
H  2.056555 -0.917010 -1.251775
\end{verbatim}
-----------------------------------------------\\
Periodic Box: ABC 15 15 15; ALPHA BETA GAMMA 90 90 90\\

\textbf{The high-spin polaron configuration on the rutile $\mathrm{TiO}_2(110)$ surface with one oxygen vacancy (multiplicity $=3$)}
\begin{verbatim}
239
 i =      105, E =     -7199.5258975861
 Ti         0.7754147603        3.8745495379       16.1782738580
 Ti         0.7712568601        0.5879588652       19.1640134650
 Ti         2.2543666625        0.6058407808       15.9815871202
 Ti         2.2670383263        3.8668341491       19.5304452883
  O         0.7507508834        0.5854703741       14.7816614182
  O         0.7570691842        3.8730662279       18.0475520568
  O         0.7564610198        0.5700611085       17.3339962637
  O         0.7767656896        3.8834544297       20.6023156245
  O         2.2379025529        5.1474664879       16.0703860398
  O         2.2398840814        1.8153484713       19.4846141310
  O         2.2403540473        2.5515901580       16.0715804531
  O         2.2602548366        5.9297782024       19.4806748549
 Ti         3.7391500301        3.8729777581       16.1801959643
 Ti         3.7355655116        0.5865379108       19.1890290597
 Ti         5.2460617696        0.6006064412       16.0695227200
 Ti         5.2314025758        3.8563634027       19.5340033304
  O         3.7285906441        0.5752391444       14.7897808310
  O         3.7303842453        3.8802606548       18.0494829778
  O         3.7169677202        0.5694337122       17.3421399565
  O         3.7405167777        3.8905401303       20.6049059781
  O         5.2120798831        5.1709822081       16.0730335134
  O         5.1317887331        1.8176165160       19.4946629489
  O         5.2067926633        2.5646633277       16.0732797471
  O         5.2390928076        5.9474058102       19.5177465086
 Ti         6.7076054494        3.8701032581       16.1826095597
 Ti         6.7128030553        0.5768916055       19.3291167168
 Ti         8.1378812954        0.5994767448       16.0690248753
 Ti         8.2161090667        3.8548136918       19.5341006002
  O         6.6910430344        0.5770168434       14.7911478588
  O         6.7001164469        3.8643923299       18.0524624385
  O         6.6956097376        0.5832190497       17.3237673433
  O         6.7168966385        3.8337372035       20.5978974906
  O         8.1771690416        5.1759350112       16.0710381596
  O         8.2919129407        1.8187372760       19.4955425492
  O         8.1850484980        2.5677766003       16.0743494222
  O         8.1898954209        5.9479978294       19.5057845676
 Ti         9.6793054806        3.8695537809       16.1802202065
 Ti         9.6886411603        0.5865438974       19.1874495868
 Ti        11.1331437595        0.6070885989       15.9806355374
 Ti        11.1802665771        3.8660477970       19.5327855754
  O         9.6521443766        0.5762247209       14.7903001017
  O         9.6735821031        3.8830471742       18.0473730152
  O         9.6776857572        0.5699027231       17.3409038892
  O         9.6899145966        3.8879970573       20.6053267595
  O        11.1497168123        5.1691859292       16.0681953199
  O        11.1817896181        1.8153604620       19.4888207018
  O        11.1463150220        2.5560492323       16.0696210133
  O        11.1535276388        5.9298275401       19.4965394254
 Ti         0.7549531323       10.4248673909       16.2941381992
 Ti         0.7639452643        7.1498060689       19.1942181594
 Ti         2.2352754318        7.1409260746       16.0412191855
 Ti         2.2041864206       10.4258636819       19.6484640134
  O         0.6840117462        7.1545587962       14.7400051597
  O         0.7543332809       10.4183161643       18.1478047338
  O         0.6903479017        7.1655576857       17.3795735440
  O         0.7581835468       10.4111377419       20.7512306520
  O         2.2272552495       11.7171633387       16.0686811511
  O         2.2428496916        8.3755830122       19.5264949010
  O         2.2259925805        9.1569888613       16.0739727631
  O         2.2188115517       12.4646538208       19.5364353081
 Ti         3.6969542743       10.4267286226       15.9638424612
 Ti         3.7496181282        7.1480515969       19.2038594495
 Ti         5.1639651653        7.1409942535       15.9957187103
 Ti         4.9821656493       10.4371674562       19.2035091761
  O         3.7783450342        7.1661905018       14.7434484154
  O         3.6380128988       10.4197774643       18.0391087929
  O         3.7861996268        7.1625892914       17.3890234973
  O         3.7601611775       10.4037095845       20.5935108339
  O         5.2083826416       11.7028222765       16.0943949102
  O         5.2313803542        8.4156557358       19.5285280314
  O         5.2115265556        9.1476900989       16.0959636304
  O         5.1351458981       12.4232513683       19.5289293748
 Ti         6.6900818475       10.4232048064       15.8935748568
 Ti         6.7151257998        7.1763438835       19.1727171905
 Ti         8.1754607578        7.1387312949       15.9954187534
 Ti         8.4092861120       10.4383828307       19.2013250260
  O         6.6916767797        7.1623921435       14.7870983220
  O         6.6979749003       10.4438623740       18.3847482420
  O         6.7096099866        7.1617993023       17.3560818361
  O         8.1730487323       11.7009000773       16.0960581127
  O         8.1906487001        8.4136844440       19.5159662362
  O         8.1685362554        9.1434597380       16.0920373497
  O         8.2803327505       12.4259878363       19.5253574259
 Ti         9.6852879476       10.4256527530       15.9639629303
 Ti         9.6782039266        7.1476474963       19.1811744257
 Ti        11.1822268958        7.1355194114       15.9948951615
 Ti        11.1858586763       10.4261698027       19.6510125670
  O         9.6454025988        7.1623488715       14.7862243185
  O         9.7553873808       10.4173377068       18.0391660827
  O         9.6486698947        7.1621412748       17.3577981272
  O         9.6321975784       10.4081618703       20.5947230263
  O        11.1515172315       11.7136928299       16.0660167896
  O        11.1643308387        8.3752909145       19.5432564568
  O        11.1505198042        9.1424349574       16.0705230723
  O        11.1924427235       12.4647018199       19.5344754578
 Ti         0.6996899724        0.5910815126        6.3920901434
 Ti         0.7011911217        3.8695047585        9.3995022235
 Ti         0.7664134571        0.5936736740       12.7912768326
 Ti         2.1815577127        3.8699083155        6.0423514170
 Ti         2.1932235033        0.5950401892        9.5796229081
 Ti         2.2718599980        3.8709309724       12.7817287158
  O         0.6990424016        3.8680958463        4.9715984127
  O         0.7062318894        0.5897380822        8.2143153976
  O         0.7203580495        3.8678044130       11.4942009032
  O         0.7043533542        3.8699295844        7.5200142118
  O         0.7122414414        0.5801393222       10.7694129411
  O         0.7288518309        3.8682376416       14.0489959439
  O         2.1849507356        1.8095672946        6.0767790971
  O         2.1913660620        5.1778995125        9.4920795993
  O         2.2205585811        1.8383513682       12.7790695940
  O         2.1843156610        5.9293715874        6.0715738946
  O         2.1916097292        2.5593021957        9.4928264982
  O         2.2108307801        5.9001262998       12.7737629536
 Ti         3.6697340524        0.5910248005        6.3906513656
 Ti         3.6714983233        3.8696498076        9.3994746839
 Ti         3.7402523696        0.5891679801       12.8156512295
 Ti         5.1518691935        3.8697714628        6.0429879022
 Ti         5.1605057784        0.5945082734        9.5832217049
 Ti         5.2421993995        3.8721547803       12.7833463400
  O         3.6691522153        3.8681908253        4.9716472548
  O         3.6764458374        0.5893664842        8.2139571665
  O         3.6889113324        3.8674015571       11.4942439140
  O         3.6743818482        3.8697334085        7.5199278853
  O         3.6817893354        0.5805672514       10.7694205697
  O         3.7036130229        3.8703245007       14.0483931233
  O         5.1545111139        1.8096835213        6.0767945720
  O         5.1616998426        5.1780819978        9.4921321252
  O         5.1917887439        1.8363210252       12.7786684985
  O         5.1545886543        5.9292536968        6.0712785860
  O         5.1616824210        2.5599282544        9.4927768056
  O         5.1913788429        5.8983852920       12.7595586507
 Ti         6.6396196320        0.5911361370        6.3913876443
 Ti         6.6410639851        3.8688163331        9.4004552539
 Ti         6.7064462916        0.5893927644       12.8109644162
 Ti         8.1212979523        3.8696696972        6.0433610208
 Ti         8.1302808899        0.5945866964        9.5829268653
 Ti         8.2106343026        3.8706483362       12.7826957772
  O         6.6391976566        3.8679669745        4.9715463993
  O         6.6456855801        0.5895719754        8.2132585578
  O         6.6611661018        3.8647996494       11.4937466201
  O         6.6442827798        3.8696966226        7.5195185747
  O         6.6519563719        0.5802878204       10.7700260573
  O         6.6715402276        3.8681639752       14.0515897350
  O         8.1248718468        1.8096213043        6.0770604476
  O         8.1313157859        5.1780525016        9.4910066459
  O         8.1578666137        1.8364394233       12.7796388590
  O         8.1242445131        5.9293173149        6.0718576590
  O         8.1316036788        2.5595223087        9.4911483026
  O         8.1479862641        5.9008624123       12.7715198880
 Ti         9.6097257847        0.5910247737        6.3908034068
 Ti         9.6105270759        3.8691133007        9.4000801061
 Ti         9.6733885744        0.5895276965       12.8173225969
 Ti        11.0911010247        3.8697895344        6.0425810290
 Ti        11.0971703751        0.5949158839        9.5805819583
 Ti        11.1799104077        3.8718670472       12.7817632201
  O         9.6088464713        3.8680983288        4.9716304186
  O         9.6168033578        0.5894496848        8.2139554608
  O         9.6265524678        3.8639947957       11.4931674196
  O         9.6149269775        3.8694549966        7.5197421053
  O         9.6209646685        0.5805233465       10.7692609135
  O         9.6415303962        3.8709567852       14.0497972885
  O        11.0945662322        1.8095213776        6.0765652855
  O        11.1005297275        5.1783082101        9.4928997079
  O        11.1289386310        1.8378362235       12.7785557498
  O        11.0938963538        5.9294063889        6.0712386433
  O        11.1023403367        2.5592876561        9.4932216304
  O        11.1075994870        5.8975095052       12.7563070295
 Ti         0.6988508766        7.1482961558        6.3858723332
 Ti         0.7016604296       10.4247421483        9.3274124798
 Ti         0.7094358559        7.1453088570       12.8144939747
 Ti         2.1810280483       10.4246763314        5.9927782093
 Ti         2.1876518981        7.1436014057        9.5797708769
 Ti         2.1704788411       10.4227022057       12.7670720454
  O         0.6989976127       10.4272471386        4.9242794160
  O         0.7028425160        7.1503426245        8.2063150396
  O         0.7283997603       10.4266947122       11.4925039747
  O         0.7038107718       10.4250374710        7.4731298197
  O         0.7070519419        7.1607253842       10.7576884210
  O         0.7319499484       10.4221708939       14.0263642043
  O         2.1840689040        8.3751214646        6.0577379694
  O         2.1927294612       11.7242759945        9.4669296707
  O         2.2136708693        8.4163823750       12.7697702198
  O         2.1846583810       12.4759317934        6.0640863220
  O         2.1924915233        9.1297943827        9.4653875712
  O         2.2204008869       12.4332684457       12.7766409069
 Ti         3.6690086499        7.1483965588        6.3853732290
 Ti         3.6690847655       10.4246148820        9.3210675696
 Ti         3.6803377831        7.1482167573       12.8112079888
 Ti         5.1507571856       10.4248028998        5.9830570324
 Ti         5.1558997883        7.1439399968        9.5777292486
 Ti         5.1540763980       10.4207699373       12.6378210318
  O         3.6666256370       10.4271588401        4.9196643306
  O         3.6750497786        7.1508566849        8.2056246082
  O         3.6955024656       10.4272138293       11.4863483626
  O         3.6757766889       10.4251329143        7.4692457146
  O         3.6786826758        7.1597270481       10.7577006991
  O         3.7132531934       10.4227441273       14.0515981719
  O         5.1544274846        8.3749976053        6.0536581904
  O         5.1637771492       11.7237350728        9.4671016735
  O         5.1914457381        8.4249386143       12.7496612023
  O         5.1543038823       12.4760311525        6.0604512999
  O         5.1637747059        9.1296642187        9.4658810676
  O         5.1895856348       12.4279931692       12.7709730760
 Ti         6.6388722908        7.1482591753        6.3849019614
 Ti         6.6418862634       10.4252944285        9.3096113623
 Ti         6.6502083797        7.1485623984       12.7996663667
 Ti         8.1234856771       10.4248919681        5.9837011745
 Ti         8.1272763980        7.1437379943        9.5741979806
 Ti         8.1956072080       10.4228215672       12.6446649796
  O         6.6392535159       10.4273418093        4.9163491238
  O         6.6436225019        7.1506631869        8.2059431651
  O         6.6690111512       10.4300453562       11.4874706625
  O         6.6433754216       10.4251981306        7.4640449759
  O         6.6501724117        7.1599004024       10.7602508048
  O         6.6749370642       10.4253021110       14.0639523262
  O         8.1242111780        8.3750181961        6.0544966534
  O         8.1308874319       11.7241005021        9.4646847538
  O         8.1490768747        8.4231531356       12.7642457268
  O         8.1247608892       12.4759841370        6.0610251150
  O         8.1304640743        9.1296709491        9.4637830083
  O         8.1565310155       12.4284044030       12.7727698991
 Ti         9.6090605468        7.1483983807        6.3858902979
 Ti         9.6138655091       10.4249248376        9.3222357011
 Ti         9.6220963041        7.1488583203       12.7966811080
 Ti        11.0923755928       10.4247785145        5.9931435392
 Ti        11.1001058741        7.1435421580        9.5787064163
 Ti        11.1821120554       10.4217401758       12.7658477038
  O         9.6112489893       10.4272488133        4.9197238507
  O         9.6141388757        7.1507871310        8.2068248354
  O         9.6336072519       10.4300510946       11.4834732177
  O         9.6125333974       10.4253271341        7.4690911168
  O         9.6136653815        7.1601212920       10.7603740708
  O         9.6377007616       10.4239529563       14.0526732589
  O        11.0939969803        8.3750474315        6.0572593219
  O        11.1027796159       11.7243243596        9.4673286724
  O        11.1063600988        8.4189898491       12.7538630916
  O        11.0945706429       12.4759903721        6.0637901322
  O        11.1011359998        9.1294355137        9.4661071600
  O        11.1257144346       12.4334967479       12.7750217146
\end{verbatim}
-----------------------------------------------\\
Periodic Box: ABC 11.8800 13.1107 25.6630; ALPHA BETA GAMMA 90 90 90\\

\textbf{The rutile $\mathrm{TiO}_2(110)$ surface with one adsorbed $\mathrm{OH}$ group (multiplicity $=2$)}
\begin{verbatim}
     242
 i =      158, E =     -7232.1207899123
 Ti        -0.6945185052       -0.4986588541       15.4663068476
 Ti        -0.6965269600       -3.7785533355       18.4649553839
 Ti        -2.1811542761       -3.7780743415       15.2832756005
 Ti        -2.1817641052       -0.4927259742       18.8268848616
  O        -0.6965957298       -3.7763854957       14.0815319891
  O        -0.7009146076       -0.5092947061       17.3368909144
  O        -0.6960184302       -3.7779023224       16.6358559495
  O        -0.6990185483       -0.5151301526       19.8934424549
  O        -2.1824754767        0.8011728960       15.3589602591
  O        -2.1813672083       -2.5590399695       18.7741832862
  O        -2.1829513259       -1.8113250017       15.3587017424
  O        -2.1886147430        1.5502569074       18.7709510861
 Ti         2.2739380545       -0.5016353527       15.4718259204
 Ti         2.2739460648       -3.7770714612       18.4639443732
 Ti         0.7886673435       -3.7780541364       15.2829863186
 Ti         0.7857531417       -0.4908536296       18.8232480765
  O         2.2739292289       -3.7761472170       14.0816593698
  O         2.2739236412       -0.5181335656       17.3379744948
  O         2.2739308267       -3.7773149724       16.6349048842
  O         2.2739035101       -0.4902363310       19.8870423693
  O         0.7849022061        0.7924814748       15.3709041757
  O         0.7887836695       -2.5582342410       18.7780944855
  O         0.7873643790       -1.8123533556       15.3601507874
  O         0.7515247321        1.5405688538       18.7700507182
 Ti         5.2423864460       -0.4986449230       15.4663110950
 Ti         5.2443894960       -3.7785502778       18.4649609141
 Ti         3.7591989008       -3.7780499529       15.2829833353
 Ti         3.7620282613       -0.4908533589       18.8232478196
  O         5.2444527143       -3.7763851551       14.0815311079
  O         5.2487722987       -0.5092827721       17.3368898525
  O         5.2438771280       -3.7779010112       16.6358594783
  O         5.2468258212       -0.5151057809       19.8934293250
  O         3.7629267973        0.7924845619       15.3709825620
  O         3.7590313111       -2.5582093875       18.7780872500
  O         3.7604882779       -1.8123548532       15.3601456081
  O         3.7962598378        1.5405978102       18.7700892604
 Ti         8.2139274575       -0.5018273260       15.4675450028
 Ti         8.2139405002       -3.7791478906       18.4625327474
 Ti         6.7290166731       -3.7780723233       15.2832802936
 Ti         6.7295597181       -0.4927242316       18.8268876040
  O         8.2139288640       -3.7764590779       14.0807895950
  O         8.2139225547       -0.5073951059       17.3363928293
  O         8.2139286901       -3.7774884961       16.6339476355
  O         8.2139104886       -0.5093108745       19.8951962362
  O         6.7303064852        0.8011684808       15.3589540382
  O         6.7291719115       -2.5590367179       18.7741802778
  O         6.7308062044       -1.8113286420       15.3587044741
  O         6.7364460693        1.5502674244       18.7709791740
 Ti        -0.6949936730        6.0537808006       15.4627955134
 Ti        -0.6832717838        2.7757555586       18.4555975981
 Ti        -2.1669682083        2.7763159408       15.2643565148
 Ti        -2.1820373344        6.0460521972       18.8268756427
  O        -0.6903542881        2.7794303342       14.0924468985
  O        -0.7014437832        6.0682797444       17.3364256866
  O        -0.7112506806        2.7756738797       16.6290468209
  O        -0.6940299379        6.0639964250       19.8910473151
  O        -2.1838249280        7.3686756748       15.3599532087
  O        -2.1866552811        4.0022904170       18.7734935658
  O        -2.1818935804        4.7565923462       15.3593238197
  O        -2.1814388775        8.1104085676       18.7761496755
 Ti         2.2739436317        6.0576788982       15.4661648225
 Ti         2.2739358121        2.7833290929       19.0387992276
 Ti         0.8554487727        2.7769506980       15.4124990631
 Ti         0.7839326573        6.0472410385       18.8101291146
  O         2.2739593215        2.7785747870       14.1029798669
  O         2.2739288429        6.0839607058       17.3360179005
  O         2.2738198734        2.7805577848       16.6324125670
  O         2.2739015330        6.0095447966       19.8896896134
  O         0.7869649958        7.3707132721       15.3595177679
  O         0.7465487075        4.0195704300       18.7429761446
  O         0.7847107752        4.7662002057       15.3708621157
  O         0.7914678480        8.1100815326       18.7823974817
 Ti         5.2428672739        6.0537781981       15.4628303750
 Ti         5.2311482992        2.7757948285       18.4557048607
 Ti         3.6921921668        2.7769521348       15.4129477553
 Ti         3.7638236363        6.0472260888       18.8101956181
  O         5.2381477363        2.7794295102       14.0924626058
  O         5.2493425083        6.0683195319       17.3364186877
  O         5.2591152676        2.7756529981       16.6291148455
  O         5.2418637313        6.0640008483       19.8910520035
  O         3.7608962964        7.3707123242       15.3595055820
  O         3.8013149964        4.0196189113       18.7428653468
  O         3.7631389935        4.7661969058       15.3709456915
  O         3.7563568963        8.1100854293       18.7824105044
 Ti         8.2139330756        6.0562415052       15.4672267745
 Ti         8.2139324553        2.7765140315       18.4427031454
 Ti         6.7146551983        2.7763166856       15.2642581563
 Ti         6.7298192487        6.0460794943       18.8268797331
  O         8.2139064096        2.7787900528       14.0770954030
  O         8.2139377930        6.0600047676       17.3389214866
  O         8.2139205043        2.7790817146       16.6221538431
  O         8.2139139504        6.0613347060       19.8962297552
  O         6.7316965833        7.3686828077       15.3599496384
  O         6.7344703188        4.0022986787       18.7734947228
  O         6.7297336603        4.7565911492       15.3593271324
  O         6.7292546158        8.1104199245       18.7761523975
  H        14.1528754850       16.8727278867       21.2730304964
  O        14.1539477010       16.0041429912       20.8290924144
 Ti        -0.6959716529       -3.7728351980        5.7062657863
 Ti        -0.6958728012       -0.4306361708        8.7060426051
 Ti        -2.1811276150       -0.4938057644        5.3581280866
 Ti        -2.1811372184       -3.6944974004        8.8806223354
  O        -0.6955928324       -0.4900402792        4.2885507772
  O        -0.6960381008       -3.7597774775        7.5297963724
  O        -0.6959265358       -0.4889625739        6.8336961165
  O        -0.6960058539       -3.7597077275       10.0769566792
  O        -2.1810609595       -2.5492925997        5.3905606266
  O        -2.1814565397        0.8252724222        8.8002176201
  O        -2.1803204289        1.5632222375        5.4089594421
  O        -2.1809301519       -1.7876928535        8.8024576931
 Ti         2.2739061706       -3.7728697603        5.7064565641
 Ti         2.2739256684       -0.4298267538        8.7063399347
 Ti         0.7888302131       -0.4934602585        5.3597730544
 Ti         0.7889880678       -3.6948944272        8.8806006618
  O         2.2738822809       -0.4896109018        4.2895604737
  O         2.2739097188       -3.7597215197        7.5300792131
  O         2.2738916983       -0.4887300036        6.8345641176
  O         2.2739132309       -3.7596840076       10.0771786149
  O         0.7889826747       -2.5492736115        5.3912259069
  O         0.7889423316        0.8252665141        8.8061703805
  O         0.7888612225        1.5632352714        5.4133383602
  O         0.7888853277       -1.7876356828        8.8029435383
 Ti         5.2437784969       -3.7728350961        5.7062713202
 Ti         5.2437203173       -0.4306399097        8.7060414571
 Ti         3.7589389218       -0.4934607290        5.3597732906
 Ti         3.7588227441       -3.6948969283        8.8806029062
  O         5.2433542270       -0.4900390079        4.2885566982
  O         5.2438530853       -3.7597763403        7.5297995674
  O         5.2437103786       -0.4889633788        6.8337023123
  O         5.2438307636       -3.7597082415       10.0769571304
  O         3.7587847193       -2.5492737936        5.3912279216
  O         3.7588750199        0.8252656387        8.8061771084
  O         3.7589163708        1.5632335773        5.4133472092
  O         3.7589298108       -1.7876339828        8.8029489690
 Ti         8.2139037834       -3.7728361629        5.7059836962
 Ti         8.2139206949       -0.4299044857        8.7047007915
 Ti         6.7289020881       -0.4938061192        5.3581303346
 Ti         6.7289476933       -3.6944957450        8.8806228528
  O         8.2138793956       -0.4900139130        4.2875565534
  O         8.2139059029       -3.7598192713        7.5295935402
  O         8.2138909045       -0.4893666445        6.8321929335
  O         8.2139114756       -3.7596692747       10.0766346873
  O         6.7288253137       -2.5492941022        5.3905633743
  O         6.7292718096        0.8252738798        8.8002230611
  O         6.7280931750        1.5632245474        5.4089648234
  O         6.7287421337       -1.7876948542        8.8024598491
 Ti        -0.6947638176        2.7825903209        5.7303452163
 Ti        -0.6950519983        6.1239370082        8.7042587812
 Ti        -2.1811354266        6.0559146620        5.3563747548
 Ti        -2.1737560531        2.8619760037        8.8977263669
  O        -0.6956625587        6.0584165881        4.2865464504
  O        -0.6934030865        2.7958948996        7.5499929955
  O        -0.6960033766        6.0716311514        6.8321590775
  O        -0.7003509335        2.7978562428       10.0958558510
  O        -2.1804101363        4.0064522765        5.4051992903
  O        -2.1809280533        7.3842045560        8.7976742559
  O        -2.1810435343        8.1185422594        5.3926286177
  O        -2.1818624841        4.7722428130        8.8054455695
 Ti         2.2739050835        2.7825507862        5.7392831660
 Ti         2.2739255939        6.1211807491        8.7042048906
 Ti         0.7888040750        6.0554635664        5.3577028087
 Ti         0.7890204039        2.8604876284        8.9284642263
  O         2.2738823896        6.0580914826        4.2873624599
  O         2.2739032956        2.7957773820        7.5559583469
  O         2.2738928287        6.0713426432        6.8327653507
  O         2.2739066208        2.7978655289       10.1051418195
  O         0.7888623224        4.0063350759        5.4088163164
  O         0.7889076562        7.3842854521        8.7979783477
  O         0.7889621979        8.1184264754        5.3933556842
  O         0.7889785804        4.7734543568        8.8119389389
 Ti         5.2425644575        2.7825910633        5.7303566952
 Ti         5.2429006192        6.1239320872        8.7042669115
 Ti         3.7589648548        6.0554643830        5.3577109228
 Ti         3.7588001721        2.8604865852        8.9284852338
  O         5.2434230194        6.0584184697        4.2865558003
  O         5.2412040197        2.7958954332        7.5500019159
  O         5.2437908557        6.0716289857        6.8321658632
  O         5.2481694305        2.7978601795       10.0958667513
  O         3.7589125835        4.0063372506        5.4088234988
  O         3.7589115583        7.3842861038        8.7979817347
  O         3.7588049868        8.1184260641        5.3933603947
  O         3.7588442360        4.7734557993        8.8119452274
 Ti         8.2138995531        2.7825204424        5.7258098628
 Ti         8.2139175412        6.1245625674        8.7031792563
 Ti         6.7289055523        6.0559149368        5.3563822631
 Ti         6.7215519773        2.8619693113        8.8977624261
  O         8.2138788447        6.0582770473        4.2857787618
  O         8.2138944122        2.7955122789        7.5436391212
  O         8.2138944043        6.0725527329        6.8308642045
  O         8.2139126341        2.7956789817       10.0861386919
  O         6.7281812919        4.0064508673        5.4052040254
  O         6.7287473037        7.3842050549        8.7976750189
  O         6.7288068895        8.1185430818        5.3926346530
  O         6.7296877652        4.7722452819        8.8054514538
 Ti        -0.6964543884       -3.7753972919       12.0804153249
 Ti        -2.1811794151       -0.4953593893       12.0793897642
  O        -0.6957429503       -0.4920889473       10.8009450249
  O        -0.6969765320       -0.5012384794       13.3526646923
  O        -2.1809900089       -2.5176618089       12.0785685172
  O        -2.1824219893        1.5247518357       12.0832447689
 Ti         2.2738528011       -3.7753438878       12.0807136290
 Ti         0.7892225345       -0.4956169359       12.0809735394
  O         2.2739360207       -0.4926077397       10.8020970415
  O         2.2739491369       -0.5005246967       13.3540683632
  O         0.7889618898       -2.5175494312       12.0797088086
  O         0.7900763892        1.5250121714       12.0884557717
 Ti         5.2441720869       -3.7753975784       12.0804144408
 Ti         3.7585858676       -0.4956176883       12.0809782816
  O         5.2436208939       -0.4920896953       10.8009452967
  O         5.2448552819       -0.5012290519       13.3526689534
  O         3.7589353578       -2.5175529310       12.0797098556
  O         3.7577903488        1.5250140163       12.0884399452
 Ti         8.2138496448       -3.7754596396       12.0809289816
 Ti         6.7289828905       -0.4953576109       12.0793819807
  O         8.2139337954       -0.4922206684       10.7994197454
  O         8.2139413424       -0.5009374029       13.3516498359
  O         6.7288790435       -2.5176573442       12.0785716488
  O         6.7302872909        1.5247457874       12.0832486406
 Ti        -0.6928815792        2.7800187061       12.1525158274
 Ti        -2.1815633320        6.0543040434       12.0794841004
  O        -0.6956737200        6.0658221133       10.8028120955
  O        -0.6971663628        6.0645884079       13.3544578256
  O        -2.1828406782        4.0411097231       12.0874653781
  O        -2.1810314465        8.0832527016       12.0753746217
 Ti         2.2738743576        2.7797822149       12.1492153009
 Ti         0.7888930306        6.0548464874       12.0801825640
  O         2.2739454672        6.0665655035       10.8043468016
  O         2.2739616133        6.0641544442       13.3562711079
  O         0.7901709093        4.0411799589       12.0933919921
  O         0.7889585976        8.0832015207       12.0765345767
 Ti         5.2405967711        2.7800187703       12.1525983704
 Ti         3.7588795953        6.0548485368       12.0801883706
  O         5.2435753161        6.0658235411       10.8028035363
  O         5.2450738465        6.0645797704       13.3544455435
  O         3.7576972941        4.0411780691       12.0933757076
  O         3.7589357422        8.0832041837       12.0765329106
 Ti         8.2138513736        2.7795941275       12.0800504639
 Ti         6.7293241069        6.0543030716       12.0794883112
  O         8.2139538999        6.0660667574       10.8010483843
  O         8.2139516297        6.0637181800       13.3534316006
  O         6.7307114525        4.0411165822       12.0874670280
  O         6.7289200733        8.0832465116       12.0753741910
\end{verbatim}
-----------------------------------------------\\
Periodic Box: ABC 11.8800 13.1107 25.6630; ALPHA BETA GAMMA 90 90 90\\

\textbf{Antiferromagnetic $\mathrm{NiO}$ (multiplicity $=1$)}
\begin{verbatim}
128

 Ni_b      0.000000000    0.000000000    0.000000000
 O      2.947503907    1.701742174    1.203313431
 Ni_b      2.947503907    0.000000000    0.000000000
 O      5.895007813    1.701742174    1.203313431
 Ni_b      5.895008000    0.000000000    0.000000000
 O      8.842511907    1.701742174    1.203313431
 Ni_b      8.842511907    0.000000000    0.000000000
 O     11.790015813    1.701742174    1.203313431
 Ni_b      1.473752000    2.552613342    0.000000000
 O      4.421255907    4.254355516    1.203313431
 Ni_b      4.421255907    2.552613342    0.000000000
 O      7.368759813    4.254355516    1.203313431
 Ni_b      7.368760000    2.552613342    0.000000000
 O     10.316263907    4.254355516    1.203313431
 Ni_b     10.316263907    2.552613342    0.000000000
 O     13.263767813    4.254355516    1.203313431
 Ni_b      2.947504000    5.105226684    0.000000000
 O      5.895007907    6.806968858    1.203313431
 Ni_b      5.895007907    5.105226684    0.000000000
 O      8.842511813    6.806968858    1.203313431
 Ni_b      8.842512000    5.105226684    0.000000000
 O     11.790015907    6.806968858    1.203313431
 Ni_b     11.790015907    5.105226684    0.000000000
 O     14.737519813    6.806968858    1.203313431
 Ni_b      4.421256000    7.657840025    0.000000000
 O      7.368759907    9.359582199    1.203313431
 Ni_b      7.368759907    7.657840025    0.000000000
 O     10.316263813    9.359582199    1.203313431
 Ni_b     10.316264000    7.657840025    0.000000000
 O     13.263767907    9.359582199    1.203313431
 Ni_b     13.263767907    7.657840025    0.000000000
 O     16.211271813    9.359582199    1.203313431
 O      4.421255907    2.552613288    3.609940369
 O      7.368759813    2.552613288    3.609940369
 O     10.316263907    2.552613288    3.609940369
 O     13.263767813    2.552613288    3.609940369
 O      5.895007907    5.105226630    3.609940369
 O      8.842511813    5.105226630    3.609940369
 O     11.790015907    5.105226630    3.609940369
 O     14.737519813    5.105226630    3.609940369
 O      7.368759907    7.657839971    3.609940369
 O     10.316263813    7.657839971    3.609940369
 O     13.263767907    7.657839971    3.609940369
 O     16.211271813    7.657839971    3.609940369
 O      8.842511907   10.210453313    3.609940369
 O     11.790015813   10.210453313    3.609940369
 O     14.737519907   10.210453313    3.609940369
 O     17.685023813   10.210453313    3.609940369
 Ni_b      2.947504000    1.701742228    4.813253877
 O      5.895007907    3.403484402    6.016567308
 Ni_b      5.895007907    1.701742228    4.813253877
 O      8.842511813    3.403484402    6.016567308
 Ni_b      8.842512000    1.701742228    4.813253877
 O     11.790015907    3.403484402    6.016567308
 Ni_b     11.790015907    1.701742228    4.813253877
 O     14.737519813    3.403484402    6.016567308
 Ni_b      4.421256000    4.254355570    4.813253877
 O      7.368759907    5.956097744    6.016567308
 Ni_b      7.368759907    4.254355570    4.813253877
 O     10.316263813    5.956097744    6.016567308
 Ni_b     10.316264000    4.254355570    4.813253877
 O     13.263767907    5.956097744    6.016567308
 Ni_b     13.263767907    4.254355570    4.813253877
 O     16.211271813    5.956097744    6.016567308
 Ni_b      5.895008000    6.806968911    4.813253877
 O      8.842511907    8.508711085    6.016567308
 Ni_b      8.842511907    6.806968911    4.813253877
 O     11.790015813    8.508711085    6.016567308
 Ni_b     11.790016000    6.806968911    4.813253877
 O     14.737519907    8.508711085    6.016567308
 Ni_b     14.737519907    6.806968911    4.813253877
 O     17.685023813    8.508711085    6.016567308
 Ni_b      7.368760000    9.359582253    4.813253877
 O     10.316263907   11.061324427    6.016567308
 Ni_b     10.316263907    9.359582253    4.813253877
 O     13.263767813   11.061324427    6.016567308
 Ni_b     13.263768000    9.359582253    4.813253877
 O     16.211271907   11.061324427    6.016567308
 Ni_b     16.211271907    9.359582253    4.813253877
 O     19.158775813   11.061324427    6.016567308
 O      7.368759907    4.254355516    8.423194246
 O     10.316263813    4.254355516    8.423194246
 O     13.263767907    4.254355516    8.423194246
 O     16.211271813    4.254355516    8.423194246
 O      8.842511907    6.806968858    8.423194246
 O     11.790015813    6.806968858    8.423194246
 O     14.737519907    6.806968858    8.423194246
 O     17.685023813    6.806968858    8.423194246
 O     10.316263907    9.359582199    8.423194246
 O     13.263767813    9.359582199    8.423194246
 O     16.211271907    9.359582199    8.423194246
 O     19.158775813    9.359582199    8.423194246
 O     11.790015907   11.912195541    8.423194246
 O     14.737519813   11.912195541    8.423194246
 O     17.685023907   11.912195541    8.423194246
 O     20.632527813   11.912195541    8.423194246
 Ni_a     1.473752000    0.850871114    2.406626938
 Ni_a     4.421255907    0.850871114    2.406626938
 Ni_a     7.368760000    0.850871114    2.406626938
 Ni_a    10.316263907    0.850871114    2.406626938
 Ni_a     2.947504000    3.403484456    2.406626938
 Ni_a     5.895007907    3.403484456    2.406626938
 Ni_a     8.842512000    3.403484456    2.406626938
 Ni_a    11.790015907    3.403484456    2.406626938
 Ni_a     4.421256000    5.956097797    2.406626938
 Ni_a     7.368759907    5.956097797    2.406626938
 Ni_a    10.316264000    5.956097797    2.406626938
 Ni_a    13.263767907    5.956097797    2.406626938
 Ni_a     5.895008000    8.508711139    2.406626938
 Ni_a     8.842511907    8.508711139    2.406626938
 Ni_a    11.790016000    8.508711139    2.406626938
 Ni_a    14.737519907    8.508711139    2.406626938
 Ni_a     4.421256000    2.552613342    7.219880815
 Ni_a     7.368759907    2.552613342    7.219880815
 Ni_a    10.316264000    2.552613342    7.219880815
 Ni_a    13.263767907    2.552613342    7.219880815
 Ni_a     5.895008000    5.105226684    7.219880815
 Ni_a     8.842511907    5.105226684    7.219880815
 Ni_a    11.790016000    5.105226684    7.219880815
 Ni_a    14.737519907    5.105226684    7.219880815
 Ni_a     7.368760000    7.657840025    7.219880815
 Ni_a    10.316263907    7.657840025    7.219880815
 Ni_a    13.263768000    7.657840025    7.219880815
 Ni_a    16.211271907    7.657840025    7.219880815
 Ni_a     8.842512000   10.210453367    7.219880815
 Ni_a    11.790015907   10.210453367    7.219880815
 Ni_a    14.737520000   10.210453367    7.219880815
 Ni_a    17.685023907   10.210453367    7.219880815
\end{verbatim}
-----------------------------------------------\\
Periodic Box: ABC 11.790016 11.790016 11.790016; ALPHA BETA GAMMA 60 60 60\\

\textbf{Antiferromagnetic $\mathrm{CoO}$ (multiplicity $=1$)}
\begin{verbatim}
128

Co_a    0.000000000    0.000000000    0.000000000
O      3.003407696    1.734018242    1.226136058
Co_a    3.003407696    0.000000000   -0.000000000
O      6.006815393    1.734018242    1.226136058
Co_a    6.006815393    0.000000000    0.000000000
O      9.010223089    1.734018242    1.226136058
Co_a    9.010223089    0.000000000   -0.000000000
O     12.013630786    1.734018242    1.226136058
Co_a    1.501703848    2.601027363    0.000000000
O      4.505111545    4.335045605    1.226136058
Co_a    4.505111545    2.601027363    0.000000000
O      7.508519241    4.335045605    1.226136058
Co_a    7.508519241    2.601027363    0.000000000
O     10.511926937    4.335045605    1.226136058
Co_a   10.511926937    2.601027363   -0.000000000
O     13.515334634    4.335045605    1.226136058
Co_a    3.003407696    5.202054726    0.000000000
O      6.006815393    6.936072968    1.226136058
Co_a    6.006815393    5.202054726    0.000000000
O      9.010223089    6.936072968    1.226136058
Co_a    9.010223089    5.202054726    0.000000000
O     12.013630786    6.936072968    1.226136058
Co_a   12.013630786    5.202054726   -0.000000000
O     15.017038482    6.936072968    1.226136058
Co_a    4.505111545    7.803082089    0.000000000
O      7.508519241    9.537100331    1.226136058
Co_a    7.508519241    7.803082089   -0.000000000
O     10.511926937    9.537100331    1.226136058
Co_a   10.511926937    7.803082089    0.000000000
O     13.515334634    9.537100331    1.226136058
Co_a   13.515334634    7.803082089   -0.000000000
O     16.518742330    9.537100331    1.226136058
Co_b     1.501703848    0.867009121    2.452272115
O      4.505111545    2.601027363    3.678408173
Co_b     4.505111545    0.867009121    2.452272115
O      7.508519241    2.601027363    3.678408173
Co_b     7.508519241    0.867009121    2.452272115
O     10.511926937    2.601027363    3.678408173
Co_b    10.511926937    0.867009121    2.452272115
O     13.515334634    2.601027363    3.678408173
Co_b     3.003407696    3.468036484    2.452272115
O      6.006815393    5.202054726    3.678408173
Co_b     6.006815393    3.468036484    2.452272115
O      9.010223089    5.202054726    3.678408173
Co_b     9.010223089    3.468036484    2.452272115
O     12.013630786    5.202054726    3.678408173
Co_b    12.013630786    3.468036484    2.452272115
O     15.017038482    5.202054726    3.678408173
Co_b     4.505111545    6.069063847    2.452272115
O      7.508519241    7.803082089    3.678408173
Co_b     7.508519241    6.069063847    2.452272115
O     10.511926937    7.803082089    3.678408173
Co_b    10.511926937    6.069063847    2.452272115
O     13.515334634    7.803082089    3.678408173
Co_b    13.515334634    6.069063847    2.452272115
O     16.518742330    7.803082089    3.678408173
Co_b     6.006815393    8.670091210    2.452272115
O      9.010223089   10.404109452    3.678408173
Co_b     9.010223089    8.670091210    2.452272115
O     12.013630786   10.404109452    3.678408173
Co_b    12.013630786    8.670091210    2.452272115
O     15.017038482   10.404109452    3.678408173
Co_b    15.017038482    8.670091210    2.452272115
O     18.020446178   10.404109452    3.678408173
Co_a    3.003407696    1.734018242    4.904544230
O      6.006815393    3.468036484    6.130680288
Co_a    6.006815393    1.734018242    4.904544230
O      9.010223089    3.468036484    6.130680288
Co_a    9.010223089    1.734018242    4.904544230
O     12.013630786    3.468036484    6.130680288
Co_a   12.013630786    1.734018242    4.904544230
O     15.017038482    3.468036484    6.130680288
Co_a    4.505111545    4.335045605    4.904544230
O      7.508519241    6.069063847    6.130680288
Co_a    7.508519241    4.335045605    4.904544230
O     10.511926937    6.069063847    6.130680288
Co_a   10.511926937    4.335045605    4.904544230
O     13.515334634    6.069063847    6.130680288
Co_a   13.515334634    4.335045605    4.904544230
O     16.518742330    6.069063847    6.130680288
Co_a    6.006815393    6.936072968    4.904544230
O      9.010223089    8.670091210    6.130680288
Co_a    9.010223089    6.936072968    4.904544230
O     12.013630786    8.670091210    6.130680288
Co_a   12.013630786    6.936072968    4.904544230
O     15.017038482    8.670091210    6.130680288
Co_a   15.017038482    6.936072968    4.904544230
O     18.020446178    8.670091210    6.130680288
Co_a    7.508519241    9.537100331    4.904544230
O     10.511926937   11.271118573    6.130680288
Co_a   10.511926937    9.537100331    4.904544230
O     13.515334634   11.271118573    6.130680288
Co_a   13.515334634    9.537100331    4.904544230
O     16.518742330   11.271118573    6.130680288
Co_a   16.518742330    9.537100331    4.904544230
O     19.522150027   11.271118573    6.130680288
Co_b     4.505111545    2.601027363    7.356816346
O      7.508519241    4.335045605    8.582952403
Co_b     7.508519241    2.601027363    7.356816346
O     10.511926937    4.335045605    8.582952403
Co_b    10.511926937    2.601027363    7.356816346
O     13.515334634    4.335045605    8.582952403
Co_b    13.515334634    2.601027363    7.356816346
O     16.518742330    4.335045605    8.582952403
Co_b     6.006815393    5.202054726    7.356816346
O      9.010223089    6.936072968    8.582952403
Co_b     9.010223089    5.202054726    7.356816346
O     12.013630786    6.936072968    8.582952403
Co_b    12.013630786    5.202054726    7.356816346
O     15.017038482    6.936072968    8.582952403
Co_b    15.017038482    5.202054726    7.356816346
O     18.020446178    6.936072968    8.582952403
Co_b     7.508519241    7.803082089    7.356816346
O     10.511926937    9.537100331    8.582952403
Co_b    10.511926937    7.803082089    7.356816346
O     13.515334634    9.537100331    8.582952403
Co_b    13.515334634    7.803082089    7.356816346
O     16.518742330    9.537100331    8.582952403
Co_b    16.518742330    7.803082089    7.356816346
O     19.522150027    9.537100331    8.582952403
Co_b     9.010223089   10.404109452    7.356816346
O     12.013630786   12.138127694    8.582952403
Co_b    12.013630786   10.404109452    7.356816346
O     15.017038482   12.138127694    8.582952403
Co_b    15.017038482   10.404109452    7.356816346
O     18.020446178   12.138127694    8.582952403
Co_b    18.020446178   10.404109452    7.356816346
O     21.023853875   12.138127694    8.582952403
\end{verbatim}
-----------------------------------------------\\
Periodic Box: ABC 12.013631 12.013631 12.013631; ALPHA BETA GAMMA 60 60 60\\

\textbf{The high-spin polaron configuration of bulk $\mathrm{CeO}_2$ containing one oxygen vacancy (multiplicity $=3$)}
\begin{verbatim}
      95
 i =       66, E =     -2250.3290837345
 Ce        -0.0400168479        0.0222497688       -0.0204723460
 Ce        -0.0364380072        2.7607324650        2.7115782356
 Ce         2.6951039370        0.0373837309        2.7071364587
 Ce         2.6979075246        2.7469309864       -0.0086989576
  O         1.3414901364        4.1377002451        4.0889542980
  O         4.0571791303        1.3882764499        1.3513094148
  O         1.3242750605        1.3931328303        1.3478325652
  O         4.0779517806        4.0813373337        4.0431647892
  O         4.0544003204        4.1219823800        1.3562216739
  O         1.3284467259        1.4160608202        4.0839169296
  O         4.0583018342        1.3860783218        4.0853357557
  O         1.3295357620        4.1318061976        1.3769248710
 Ce         5.4155353762        0.0287782414        0.0008706815
 Ce         5.4219871308        2.7291596925        2.6995820773
 Ce         8.1562575589        0.0137752378        2.7189963728
 Ce         8.1572542597        2.7532366193       -0.0276074156
  O         6.8239578898        4.0802048883        4.0529214431
  O         9.5244455715        1.3900055738        1.3483090612
  O         6.7918993373        1.3843934635        1.3469719285
  O         9.5451142671        4.1261321352        4.0780514069
  O         9.5262483870        4.1229621150        1.3422568516
  O         6.7900069157        1.3566322080        4.0838657084
  O         9.5273928302        1.3894426684        4.0798785656
  O         6.7925667927        4.1168892000        1.3356649559
 Ce        -0.0523187493        5.4989986086       -0.0064841130
 Ce        -0.0547871846        8.2222361541        2.7101498624
 Ce         2.6855584917        5.4843469939        2.7379769636
 Ce         2.6847947250        8.2257813085        0.0008898277
  O         1.2847151436        9.6359204934        4.0599488078
  O         4.0576427117        6.8553247137        1.4465542443
  O         1.3252739126        6.8562427452        1.3613440068
  O         4.0657112796        9.6272508288        4.0472481545
  O         4.0606018549        9.5877780729        1.3677596049
  O         1.2866476907        6.8559832722        4.0419181245
  O         4.0847532340        6.8225529648        4.3201509329
  O         1.3230493098        9.5943505728        1.3475917798
 Ce         5.4220357081        5.4757764959       -0.0044811296
 Ce         5.4316770754        8.2233373110        2.7416887723
 Ce         8.1776800501        5.4892462852        2.6976418664
 Ce         8.1745374689        8.2133878852        0.0065304032
  O         6.7825261551        9.5618547147        4.1035560989
  O         9.5269880548        6.8570002014        1.3543329792
  O         6.7902982561        6.8573079737        1.3633670143
  O         9.5034585557        9.5874922360        4.0883616394
  O         9.5255401231        9.5906656149        1.3518224189
  O         6.8271323088        6.8261727249        4.0508399270
  O         9.5238500993        6.8580630283        4.0843929800
  O         6.7868823368        9.5814999166        1.3921285234
 Ce        -0.0591465589        0.0397606491        5.4595176727
 Ce        -0.0525335044        2.7422766450        8.1757276973
 Ce         2.6859073237        0.0168762629        8.2038921446
 Ce         2.6891380325        2.7578926658        5.4455355136
  O         1.2760013968        4.0781774064        9.5980467678
  O         4.0576283891        1.3966549662        6.8177665747
  O         1.3271398559        1.3913934129        6.8136156400
  O         4.0904229724        4.0502461599        9.5997487116
  O         4.1127001737        4.2587461104        6.7725430112
  O         1.3253151708        1.3728576184        9.5464760713
  O         4.0583416006        1.3851544010        9.5489595236
  O         1.2720448255        4.0647080484        6.7802410656
 Ce         5.4316268708        0.0040296242        5.4603316143
 Ce         5.4310706983        2.7552700963        8.1923472607
 Ce         8.1658109248        0.0251108827        8.1799572022
 Ce         8.1776639832        2.7302346790        5.4521173781
  O         6.7880274244        4.1161010861        9.5333203677
  O         9.5237538537        1.3869942857        6.8148727967
  O         6.7905371690        1.3789356626        6.8153540138
  O         9.5056428134        4.1245891965        9.5410720852
  O         9.5249585461        4.1217975133        6.8108790730
  O         6.7877192003        1.4006718272        9.5454257269
  O         9.5246055409        1.3888796077        9.5473600437
  O         6.8386637705        4.0657348100        6.7985713844
 Ce        -0.0403087412        5.4868698340        5.4442223572
 Ce        -0.0464559817        8.2333745502        8.2033951488
 Ce         2.5987782260        5.4170827627        8.2605501331
 Ce         2.5958687543        8.3303885089        5.3982748590
  O         1.3242543377        9.5912528090        9.5566716768
  O         1.4102371162        6.8837448788        6.8049357077
  O         4.0231731953        9.6495572686        9.6213234635
  O         4.0402970133        9.4788901002        6.8216950692
  O         1.2539091372        6.8956661263        9.6118509491
  O         4.0559978081        6.8469388949        9.5899152779
  O         1.2663816386        9.6496177363        6.8475630730
 Ce         5.5292372134        5.3780896090        5.3591094225
 Ce         5.5095382845        8.2656287262        8.3234757279
 Ce         8.1561281361        5.4715608341        8.1900758765
 Ce         8.1502253254        8.2225629013        5.4553672917
  O         6.8387998892        9.6244415507        9.6182993148
  O         9.5067483709        6.8548255826        6.8180773238
  O         6.6445211231        6.7891378264        6.7854149984
  O         9.5400863433        9.5895707708        9.5511215900
  O         9.5227904072        9.5873444560        6.8161719362
  O         6.8644592383        6.8061399890        9.6155267631
  O         9.5229446369        6.8560428167        9.5490177925
  O         6.8480907719        9.6247035653        6.8016442137
 \end{verbatim}
-----------------------------------------------\\
Periodic Box: ABC 10.934901  10.934901  10.934901; ALPHA BETA GAMMA 90 90 90\\

\textbf{A snapshot from AIMD containing 32 liquid water molecules and one $\mathrm{Ag}^{2+}$ ion (multiplicity $=2$)}
\begin{verbatim}
      97
 i =    28398, time =    28398.000, E =      -586.7927473371
  O        -3.9264314541        4.0393010318       -1.6191104829
  H        -3.7621260516        4.7870914188       -2.3402770075
  H        -4.4325190722        3.2282519854       -1.8188821112
  O         0.6475688441        3.5855196700        0.3395277334
  H        -0.3672513235        3.6022007571        0.1126845541
  H         0.9740922556        3.0746799204       -0.4403188658
  O        -1.9912764498        2.8721993626       -0.0256081066
  H        -2.7509325968        3.1846583954       -0.6221121379
  H        -2.4329468862        2.6008101249        0.8373429395
  O        -2.3064191604       -3.3984711015       -5.3133193778
  H        -2.7833687139       -3.1114268882       -6.1886318231
  H        -1.6414973066       -4.1313183686       -5.5175741296
  O         5.0893059953       -0.4202585545        2.4212862937
  H         5.5844567102       -1.2451188293        2.2034398648
  H         4.3490986923       -0.4076539038        1.7529170109
  O         1.7047371086       -3.6938996784        0.6855775266
  H         1.4050381277       -4.6279481174        0.3119646236
  H         1.8539593851       -3.7975691946        1.6711901586
  O        -4.6518556048       -2.0656234162       -1.2368103938
  H        -4.0315898364       -2.6352260981       -1.7211288502
  H        -4.6499139172       -1.1479230481       -1.7221690490
  O         0.5071345466        4.8223708786       -3.1374301012
  H         0.0870672568        5.4654917006       -2.5066666653
  H         0.3488136483        3.9897765755       -2.5760017027
  O         4.8182528838        5.2463110333        0.5692479608
  H         4.5133270523        4.4473405128        1.1071367786
  H         5.3389058014        4.8330494314       -0.2082259721
  O         6.9996301907        1.7647614900        2.6400643957
  H         7.0304300556        2.1049337059        3.6155065895
  H         6.6790449796        0.8949264892        2.6302883191
  O         1.8562735910        2.0834979986       -4.9395242119
  H         1.2974622272        1.7505202425       -5.6360577362
  H         1.3987504681        1.9677042157       -4.0726952440
  O         0.2084954102        2.1012236548        2.6941384513
  H        -0.7052479584        1.8528449804        2.8543996893
  H         0.2495262940        2.5548406264        1.8603439212
  O         5.2161835458        0.2310535620       -2.5572014151
  H         6.1583423829        0.5245970757       -2.7594773213
  H         4.7074910488        0.1525908984       -3.4157862874
  O        -1.2587688435       -1.9214558549        0.6625836569
  H        -0.4617670441       -2.1594560689        1.0697299480
  H        -1.1573967752       -0.9883363828        0.4939462085
  O        -3.4247020671       -2.8243918560        2.0600667008
  H        -3.7456944314       -3.6263553621        1.5924333981
  H        -2.6694614742       -2.5850351545        1.5069082788
  O         3.2603537126        5.5583282620       -2.9336228434
  H         3.8967509089        4.9716823845       -3.4725892245
  H         2.4259724588        4.9947868073       -3.0698021646
  O         3.9503850196       -6.2467045826       -5.4421917041
  H         3.5766888001       -5.3205845822       -5.8287761641
  H         3.1976778952       -6.7389854480       -5.0739733009
  O        -2.0028863219        0.5699924556       -3.1444491727
  H        -1.6595299391        0.0473061555       -3.9070513536
  H        -2.1626725958        1.4399322083       -3.5397359759
  O        -0.3932174687        4.5298145703       -5.8837150969
  H        -0.0466600172        3.8957702622       -6.5137599269
  H         0.0572025961        4.5046987490       -4.9848645906
  O         1.2221475418        2.1301044865       -2.0371883131
  H         0.6752182142        1.3705436289       -1.6448235581
  H         2.1629513336        1.8572311968       -1.7894707753
  O         2.3244974749       -4.1579141355        3.1686487005
  H         2.3042132736       -3.2330673917        3.6103183306
  H         1.4582731909       -4.5770932084        3.4725524937
  O         2.4464816469        0.5100947666        1.5105091853
  H         1.6834602589        0.8037733963        1.9800350593
  H         3.1503409063        1.1888158554        1.5849852665
  O         3.8705562761        0.0325623130       -5.0534909075
  H         3.1225878331        0.7128800017       -4.9180525364
  H         4.1898377386        0.0102557149       -6.0227372083
  O         9.1732410397        0.5556399538       -0.6577551288
  H         8.7376816041        0.4946105736       -1.5543838269
  H         8.7114295905        1.4683437126       -0.5102890754
  O        -3.2678669889       -7.1204384112        5.2654470280
  H        -2.7901381182       -6.2600937267        5.3843222520
  H        -4.2101911078       -6.8005338779        5.1531181477
  O        -0.6563016669       -1.3236649501        5.2784418208
  H         0.3589632334       -1.5554073702        5.1360555368
  H        -1.1240923918       -2.1261151915        4.9390938715
  O         4.5158962721        2.7456942522        1.9740291718
  H         4.1363576135        2.9308136930        2.9289852136
  H         5.4677860294        2.4200539294        2.1588666149
  O        -3.1713848572       -3.7531044153       -2.9009873202
  H        -2.9507368999       -3.7250428527       -3.8842370654
  H        -2.2902278735       -3.5952801102       -2.5731813075
  O         2.1114862800       -2.1195879102       -4.8589285900
  H         2.4581864651       -2.8492305413       -4.2021979297
  H         2.8458305230       -1.3900081643       -4.8349674232
  O         2.4206609376       -1.1416416783       -0.8254335303
  H         2.2271679370       -0.4960441815       -0.1101634281
  H         1.6474530302       -1.4379013521       -1.3737977912
  O        -0.6868411768       -2.9521026914       -1.9906480359
  H        -0.6700830702       -2.3299961823       -2.8125734354
  H        -0.9419317769       -2.4487094610       -1.2015358705
  O         3.8855976637        1.9401234650       -1.0735517384
  H         4.1850178421        1.8427942111       -0.1229224705
  H         4.4677735498        1.3278330773       -1.6137228423
 Ag         3.4575735671        6.8103958212        9.5181216222
 \end{verbatim}
-----------------------------------------------\\
Periodic Box: ABC 9.860 9.860 9.860; ALPHA BETA GAMMA 90 90 90\\

\end{suppinfo}